# Saliva-Sensing Dental Floss: A Novel Tool for Assessing Daily Stress via On-Demand Salivary Cortisol Measurement Integrating Molecularly Imprinted Polymer and Thread Microfluidics

Atul Sharma [a,b], Nafize Ishtiaque Hossain [a,b], Ayanna Thomas [c], Sameer Sonkusale [a,b*]

[a] Nano Lab, Advanced Technology Laboratory, Tufts University, Medford, MA, USA.
[b] Department of Electrical and Computer Engineering, Tufts University, Medford, MA, USA
[c] Department of Psychology, Tufts University, Medford, MA 02155, USA
*Corresponding author email: sameer@ece.tufts.edu, sameer.sonkusale@tufts.edu

**Abstract**

Elevated stress negatively impacts health, increasing the risk of cardiovascular diseases, weak immunity, anxiety, and cognitive decline. Current lab-based methods lack real-time stress measurement, while saliva diagnostics offers a convenient and non-invasive alternative. This work proposes repurposing dental floss as a thread-analytical saliva diagnostic device. This method uses flossing to collect and transport saliva to a flexible electrochemical sensor via capillary microfluidics, where cortisol, a stress biomarker, is measured. By integrating an electropolymerized molecularly imprinted polymer specific to cortisol (cort-eMIP) on a porous-laser-engraved-graphene (PLEG) working electrode, the saliva-floss platform shows a detection range of 0.10-10000 pg mL$^{-1}$ ($R^2$ = 0.9916) with a detection limit of 0.023 pg mL$^{-1}$ for cortisol (in buffer medium). The saliva-sensing dental floss provides results within 5 minutes. The thread-based microfluidic design minimizes interference and ensures consistent repeatability when testing artificial and actual human saliva samples, yielding 98.64%–102.4% recoveries with a relative standard deviation of 5.01%, demonstrating high accuracy and precision. Validated using human saliva samples in a stress study, it showed a high correlation (r= 0.9910) with conventional ELISA assays. Combined with a wireless readout, this saliva floss offers a convenient way to monitor daily stress levels and can be extended to detect other critical salivary biomarkers.

## 1. Introduction

Stress arises as a natural physiological reaction to the pressures of demanding physical or psychological circumstances triggered by feelings of anxiety. While acute stress responses in healthy individuals can be adaptive and controllable, chronic stress, pervasive in society, is associated with a range of health issues, such as cardiovascular diseases, mental health disorders, and compromised cognitive functions [1,2]. Given these impacts, it is important for us to understand and evaluate stress levels early and often in the realm of preventive healthcare. The current gold standard to monitor chronic stress often relies on self-reporting questionnaires and psychiatric evaluations [3–5]. While valuable, these methods have limitations. Their inability to monitor stress in real-time data makes it less effective. Moreover, reliance on self-reporting and psychiatric evaluations introduces biases that can compromise the accuracy of assessments [6,7]. In contrast, non-invasive biomarkers such as cortisol provide a way to objectively quantify psychological and biological stress [8–13]. Several methods have been proposed; however, these methods rely on blood draws and access to centralized facilities, making them less practical [4,9,14,15]. Thus, a selective, sensitive, and cost-effective point-of-care stress assessment tool is of paramount interest.

Point-of-care diagnostics (POCs) have already enabled on-demand and continuous monitoring of various biomarkers such as glucose [10,16–19]. Saliva-based POCs are promising and have already shown potential for chronic stress monitoring due to ease of sampling and non-invasiveness. However, there are still challenges related to patient compliance [20,21]. First, the tests often require individuals to proactively remember to perform tests, which can be a barrier to real-time monitoring [21]. However, measuring stress levels in vulnerable populations can induce anxiety, altering results (the "white coat" effect) [22–24]. These limitations further underscore the need for a radically different approach to POC testing. For example, Wang et al.[19] suggested using a mouthguard enzymatic biosensor for non-invasive lactate detection in saliva for stress and fatigue monitoring. However, the device is bulky and somewhat invasive. In this manuscript, we posit a what-if scenario where dental floss automatically provides a salivary cortisol measurement without engaging the subject under test. We are encouraged by the established high correlation between the level of salivary metabolites and their corresponding blood concentration, primarily cortisol [9]. Moreover, making this measurement a natural part of the subjects' everyday lives may increase compliance and support routine monitoring.

Reported saliva-based POC platforms for cortisol detection involve employing enzymes or their fragments to amplify [25,26] signals, and enhance sensitivity, or employing specific aptamers [27,28] including electrochemical [9,26,29] and optical detection [28,30] methodologies. Immunoassays and immunosensors are prevalent among the described electrochemical devices, favored for their simplicity, and high sensitivity. Despite delivering satisfactory performance with sensitivity in sub ng/mL levels [31], these platforms require complex sample processing such as dilution and centrifugation, making them impractical for at-home use [25–27] Moreover, using antibodies and aptamers increases their cost and storage requirements. We propose using synthetic receptors, namely electropolymerized molecularly imprinted polymers (eMIPs), to overcome these challenges.

Molecularly imprinted polymers (MIPs) have emerged as promising alternatives to traditional biorecognition elements like antibodies or aptamers. They offer several advantages, including enhanced chemical and physical stability and a simple and tunable fabrication process [32–34]. The underlying concept of MIP technology revolves around employing molecular templates to form selective binding sites within cross-linked polymer scaffolds, thereby enabling the development of tailor-made recognition sites for target analytes. MIPs have already been developed for selective recognition of salivary cortisol [35,36]. However, existing platforms require ferrocyanide/ferricyanide redox couple as external signaling probes. Unfortunately, this reliance on external probes hinders their practicality for POC applications due to the necessity for sample and sensor preparation [37]. To address this challenge, we embed an "internal redox signaling probe" into the eMIP matrix, which generates a signal upon cortisol (template) binding.

In this work, we designed a sensing dental floss with a thread-based microfluidic channel integrated with cortisol eMIPs (cort-eMIPs) on a porous laser engrave graphene electrode (PLEG) to monitor cortisol levels in saliva. It comprises four interconnected regions: the flossing region of the thread that collects saliva, the microfluidic thread region to transport saliva, the sensing region connected to a flexible electrochemical sensor, and the waste zone. Utilizing an FDA-approved suture as a floss, saliva is collected and directed due to capillary action on the attached thread to the sensing area where a flexible cort-eMIP sensor is connected to the thread. An electropolymerized cortisol-imprinted polypyrrole-prussian blue (PPy-PB) matrix on a PLEG as a working electrode (WE), silver/silver chloride (Ag/AgCl) as a reference electrode (RE), and bare PLEG as a counter electrode (CE) make the cort-eMIP sensor. The PB acts as an internal redox signaling probe. The choice of pyrrole was guided by the ease of electropolymerization to form a PPy-PB network, which is a stable interferent-free conductive surface at physiological pH levels [37,38]. The cortisol in the saliva binds to the cort-memory sites in the cort-eMIP layer. This binding inhibits the redox conversion of the embedded PB layer, resulting in a change in the amperometric current in a chronoamperometric (CA) measurement. The captured CA signal is wirelessly transmitted to a remote device or accessed through a mobile app. The dental floss POC platform can detect salivary cortisol in the 0.10- 10000 pg mL$^{-1}$ range with the lowest detection limit (LOD) of 0.023 pg mL$^{-1}$ (in buffer). In contrast to conventional sensors to POC-based platforms for salivary cortisol detection [9,27,39–42], our saliva-sensing dental floss stands out for its on-demand, easy-to-use, and portable platform without engaging the subject under test. The floss platform demonstrates higher sensitivity due to integrating PLEG electrodes, eMIP, and threads as flossing and microfluidic channels. Additionally, the convenience of wireless CA readout enables the saliva-sensing dental floss device to effectively monitor the smallest fluctuations in salivary cortisol with exceptional accuracy and precision in real-time.

## 2. Results and Discussion

### 2.1. Design and overview of the saliva-sensing dental floss platform

A saliva-sensing dental floss was chosen for its biocompatibility, lightweight and portability, and ability to seamlessly integrate with the individual's daily lifestyle for routine stress monitoring [43]. Figures 1(a-b) and S1(a-f) present a schematic representation of the saliva-sensing dental floss designed for salivary cortisol monitoring. The designed saliva floss (base layer and top cover) was fabricated by 3D printing using FDA-approved biocompatible dental resin (Figure 1c) (see experimental section). It comprises four interconnected thread regions: (i) the flossing region made up of flossing thread designed for the collection of saliva, (ii) the microfluidic thread processed through plasma treatment (Figure 1d), facilitating saliva transportation through capillary action and filtering off the larger particles or food debris (Figure 1b, Figure S1 and Figure S2), (iii) the sensing portion interfaced directly with a flexible electrochemical sensor fabricated on PLEG electrode using eMIP methodology (Figure S3), and (iv) the waste zone made up an adsorbent pad to collect saliva after analysis (Figure S2). We used an in-house fabricated PLEG-electrode (Figure 1e and Figure S4) functionalized with the cort-eMIPs layer for cortisol quantification. Laser-engraved carbonization of thin, flexible polyimide (12.5 μm) substrate was achieved by $CO_2$ laser under optimized engraving conditions (see experimental section). Fabricated PLEG electrodes showed a high surface area and excellent electrical conductivity (32 ± 3 Ω/cm, n=10) and are amenable to chemical surface modification to realize sensors [44,45]. During practical use, saliva floss might encounter some bending or flexing during flossing. Following fabrication, PLEG electrode bending capability was evaluated when bent from 10 % to 90 %. PLEG electrodes exhibited excellent flexibility and bending capability when bent at different angles (>50 % and >90 %) (Figure h and Figure i). This flexibility is crucial due to the floss's bending nature; hence, all integrated components must possess flexible properties. Figure 1b demonstrates the utility of developed saliva floss for cortisol sensing with a wireless signal transmission capability to a mobile device.

The eMIP working electrode (cort-eMIP/PPy/PB film) was made in situ for cortisol detection, incorporating PB as an internal redox signal probe and cortisol as a template. To quantify cortisol, when the cort sensor is incubated with a sample or medium containing cortisol molecules, cortisol diffuses and binds into cortisol memory sites grafted during sensor fabrication. The cortisol in saliva was detected using the chronoamperometric method under +0.10 V, which was chosen from the oxidation of PB in cyclic voltammetric curves. The binding of cortisol molecules to the PPy-PB layer obstructs the local charge distribution. It affects the redox state of PB by impeding the electron transfer pathways and further decreases the PB oxidation process since the eMIP cavities are filled with cortisol from saliva. This amperometric signal corresponds to the amount of cortisol in the saliva, providing a direct measure of cortisol levels. The changes or fluctuations in cortisol levels can easily be correlated with individual stress levels, making this technology a promising tool for stress monitoring.

**Figure 1**

## 2.2. Preparation of cortisol imprinted sensor (cort-eMIP/PPy/PB/PLEG)

Figure 2(a-d) demonstrates the crucial step of electrochemical polymerization of cort-specific MIP film embedded with PB on the PLEG, which is key to fabricating a cort-sensor. Prior to this, the surface modification of the bare PLEG electrode was meticulously carried out in a 0.50 M sulphuric acid ($H_2SO_4$) solution through electrochemical oxidation (Figure S5). This modification method plays a pivotal role in increasing the surface roughness and hydrophilicity of the electrode surface by introducing a hydroxy group [46]. This, in turn, will help in increasing the surface adhesion and also set the foundation for the subsequent polymerization process [46–48]. After the surface of the PLEG electrode was modified, the electrochemical polymerization procedure meticulously generated a thin conductive polymer layer on the PLEG (Figure 3a). The PLEG electrode was coated entirely with a PPy/PB layer imprinted with cortisol. A clear redox peak of cort-eMIP/PPy/PB/PLEG process associated with the PB redox cycle is evident, providing cathodic and anodic peaks around 0.12 V and 0.36 V, respectively. The voltammogram (current) increases with subsequent cycles (1-20 cycles), a characteristic indicating the formation of a PPy-conducting polymeric layer. However, peak-to-peak separation becomes less with an increase in the number of cycles and stabilizes around the $20^{th}$ cycle. As the eMIP film grows onto the electrode, more and more cortisol molecules diffuse through the precursor mixture and are embedded in the resulting film. Further cycling the potential leads to the formation of thicker PPy films, which can impact the performance and sensitivity of the resulting cort-eMIP sensor. Therefore, the number of electropolymerization cycles was precisely and meticulously optimized to 20 cycles. In contrast, e-NIP fabrication (Figure 3b) exhibited a higher current during the first polymerization cycle. Still, no significant change in the current with further cycling was observed

(Figure 3b). This is plausibly due to the absence of cortisol molecules, which are electrochemically inactive and hinder charge transfer during e-MIP formation.

**Figure 2**

During the electropolymerization (eMIP fabrication), two types of interactions occur: (i) hydrogen bonding interaction between the hydroxy group of cortisol and the imide nitrogen of PPy and (ii) hydrogen bonding interaction between the carboxy group of cortisol and the imide nitrogen of PPy. However, it is essential to note that conventional template extraction through washing with organic solvents, such as methanol acetic acid, might lead to solvent-induced swelling of the PPy/PB scaffold, potentially promoting nonspecific binding or fouling [49,50]. To address this, we emphasize the importance of the over-oxidation of the PPy/PB scaffold in a PBS medium to get the cortisol molecule out of the PPy-matrix (Figure 3c), which builds the complementary recognition sites for the cortisol molecule. Figure 3c and Figure 3d depict the voltammograms of both e-MIP and e-NIP, showing a decrease in current magnitude with successive cycles. Excessive oxidation of PPy resulted in cortisol extraction from the PPy matrix, forming complementary binding sites (Figure S6). This indicates the removal of weakly bound monomers and cortisol molecules from the e-MIP film post-overoxidation process (Figure 3c-d and Figure S6). The overoxidation process turns PPy doped with negatively charged functional groups, which is reflected by the decrease in the redox signal (Figure 3c). In Figure 3d, the overoxidation of the eNIP initiates irreversible PPy/PB scaffold oxidation. We tuned the thickness of the eMIP layer by increasing the electropolymerization cycles from 5-30 cycles to explore the impact of the thickness of the MIP layer on the performance of our saliva-sensing dental floss POC platform. The highest current change was achieved at 20 cycles. Also, to ensure effective conditioning of the eMIPs film, we studied the overoxidation process by increasing the number of cycles from 1-

20 cycles. No change in current was observed after 15-16 cycles. However, to ensure that no trace of cortisol was present in the MIP layer, we chose 20 cycles of overoxidation. Additionally, during template extraction in an aqueous buffer medium, the slightly lower positive potential was used, which is due to avoiding the change in the oxidation state of the PPy scaffold and conductivity [51], which might result in a decrease or breaking of the bonding interaction created in the presence of cortisol [52].

Scanning electron microscopy (SEM) revealed that the electrodeposited cort-eMIP layer with 20 cycles exhibited the most uniform surface coverage (Figure S6d) and porous electrode surface post-overoxidation (Figure S6e). Likewise, the fabrication of the eNIP electrode applied the same methods as those used for eMIP, except the solution did not contain cortisol, which resulted in no porous surface (Figure S6f). When not used, the prepared sensor strip was stored at room temperature in a nitrogen environment.

**Figure 3**

## 2.3. Sensor characterization and surface studies

### 2.3.1. Electrochemical studies of cort-eMIP/PPy/PB sensor

Overoxidation might affect the electrochemical properties of the cort-eMIP/PPy/PB layer, which could ultimately impact the sensor's binding kinetics and sensitivity. Therefore, changes in the electrochemical activity are studied by CV (Figure 3e-f) and electrochemical impedance (EIS) (Figure 3g-h) using an external standard ferrocyanide/ferricyanide solution. The well-defined redox peak behavior and peak current of redox couple at bare PLEG electrode (curve i, Figure 3e and Figure 3f) decreases on electropolymerized MIP electrode surface (curve ii, Figure 3e and Figure 3f). This is due to the formation of the insulating eMIP and eNIP layers. However, after overoxidation of the cort-eMIP/PPy/PB in PBS, the anodic peak of the undoped eMIP layer increased from 1.20 µA to 24.82 µA (Figure 3f, curve iii). This increase is attributed to the enhanced permeability of the redox probes through the cortisol-specific cavities. Whereas rebinding of cortisol (1.0 pg mL$^{-1}$) to the cort-eMIP binding sites obstructs the redox signal at the electrode surface and causes a decrease in anodic peak current to 23.89 µA (Figure 3f, curve iv). In comparison, the eNIP electrode exhibited a lower peak current after extraction at 16.01 µA (Figure 3f, curve iii), further decreasing to 15.90 µA upon binding the same cortisol concentration (Figure 3f, curve iv). These changes in current before and after cortisol binding for eMIP (ΔI, 0.93 µA) over eNIP (ΔI, 0.11 µA) were approximately in an 8.5:1 ratio. These results are correlated with the fact that the cort-eMIP possess more cortisol binding sites and, thus, exhibited significant current than cort-eNIP. The EIS measurements (Figure 3g and Figure 3h) also aligned with the CV findings. Electropolymerization decreases the charge transfer resistance (Rct) (Figure 3g, curve ii) compared to the PLEG electrode (Figure 3g, curve i). Template removal further increased the permeability of eMIP to redox probes (Figure 3g, curve iii) and decreased the charge transfer resistance (Rct). Rebinding of cortisol is responsible for an increase in Rct (Figure 3g, curve iv) over the overoxidized electrode (Figure 3g, curve iii). A similar finding was observed with the eNIP electrode (Figure 3h, curve i- iv).

### 2.3.2. Surface studies of cort-eMIP/PPy/PB sensor

FTIR spectral analysis of eMIP (Figure 4a, green curve) shows prominent vibrational peaks around 3398 cm$^{-1}$ and 1715 cm$^{-1}$, corresponding to the -OH and -C=O stretching vibration, respectively. Additionally, a weak peak around 1214 cm$^{-1}$ is observed, which is associated with C-O stretching (next to a strong band at 1196 cm$^{-1}$) [42]. These vibrational bands are attributed to cortisol entrapment into the polymer matrix, as they are not observable in e-NIP (Figure 4a, blue curve). The primary amine peak of the PPy scaffold appears at 3346 cm$^{-1}$ for e-NIP at the shoulder of the -OH stretching peak (3398 cm$^{-1}$) for e-MIP. Other vibrational peaks at 1638, 1512, 1455, and 1408 cm$^{-1}$ are associated with C=N, C=C, and C–N stretching vibration of PPy in the PPy-PB scaffold, respectively [42,53]. The corresponding vibrational bands the template's successful incorporation into the e-MIP layer. SEM analysis reveals the surface morphology of the porous and cleaned PLEG electrode (before electropolymerization) (Figure 4b). After electropolymerization, the formation of the cort-eMIP/PPy/PB layer shows a polymeric surface morphology with small-sized porous features (Figure 4c). This porous structure is likely due to the presence of cortisol in the eMIP, which leaves pores upon elution. In contrast, the eNIP film lacks cortisol molecules (Figure 4d), enabling polymerization with more extended, branch-like formations (See supplementary information, Figure S6). Furthermore, the Raman spectrum was also characterized at different stages of eMIP formation. Figure S7 shows an increase in the $I_D/I_G$ ratio after forming the cort-eMIP ($I_D/I_G = 0.833$) scaffold on the PLEG electrode ($I_D/I_G = 0.800$). Meanwhile, the cort-eNIP electrode showed an $I_D/I_G$ of 0.805. A detailed description is provided in the supplementary section (S3).

**Figure 4**

**2.4. Performance evaluation of cort-saliva sensing dental floss platform**

**2.4.1. Effects of experimental variables**

The experimental conditions meticulously selected for fabricating PPy/PB film with and without the cortisol (as a template) significantly impact the eMIP film's growth [54,55]. For example, a higher concentration of monomer/crosslinker, the limited solubility of cortisol in an aqueous medium, and an inappropriate template-functional monomer ratio can potentially reduce the sensitivity of the eMIP layer [54,55]. In this work, we employed cyclic voltammetry (CV) to control the oxidation process, following a previously reported method [55] with slight modification (i.e., under optimized electrodeposition parameters). Firstly, the electropolymerization parameters, including the potential window and the number of cycles, were carefully adjusted (as shown in Figures 3a and 3c). This optimization ensured the formation of a uniform and highly cross-linked MIP film on the PLEG electrode, which is essential for stability. Post-overoxidation, the cort-eMIP demonstrated an oxidation peak around 0.10 V (Figure 3c) due to the oxidation of the PB-redox couple embedded into the PPy matrix. The binding capacity of the eMIP scaffold, a key factor in cortisol detection, was first evaluated in a binding buffer with and without cortisol (1.0 pg mL$^{-1}$) using CV and CA measurements (Figures 5a-b). It is clear from CV (Figure 5a) and CA (Figure 5b) that the binding of cortisol into cortisol-memory sites fabricated in the eMIP layer led to a change (decrease) in the magnitude of the oxidation peak current. At the same time, the cort-eNIP/PPy/PB sensor, which was used as a control, did not exhibit similar changes (Figure 5a-b), confirming the specificity of the eMIP layer for cortisol detection. For analytical measurement, the CA current was carried out at a fixed potential of 0.10 V across a series of eMIP sensors with increasing cortisol concentrations from 0.1 to 10000 pg mL$^{-1}$ added to a PBS medium (Figure 5c-f).

An inappropriate stochiometric combination might result in the formation of a densely packed polymeric film or inadequate complementary binding sites [50]. The effect of monomer and monomer ratio on the template was thoroughly investigated. The results from Figure S8 clearly show that a 20 mM pyrrole (as functional monomer) concentration is sufficient to electropolymerize on the electrode surface, achieving maximum surface coverage (Figure 4c and Figure S6d). No further increase in current response was observed with a further increase in pyrrole concentration, indicating the saturation of binding sites. Figure S9 demonstrates that the 1:4 molar ratio (template: monomer) is optimal for eMIP preparation, after which no more significant change in response was observed. This is due to the formation and alignment of the PPy-PB network with cortisol (template) molecules showing maximum interaction. These findings align with the existing literature, demonstrating that the correct monomer-template ratio significantly helps MIP film stability in a noncovalent eMIP approach [48,50].

Since the sensor is placed along the thread-based microfluidic channel, the total sensing volume presented to the sensor at any time is a function of the flow rate and the dimension of the sensor. This flow rate is a constant for a given thread with a fixed diameter (our prior work showed a tight control of the volumetric flow rate in the thread-based microfluidic channel) [56]. This ensures that the sample volume presented to the sensor is constant irrespective of the total saliva volume above a minimum. Constant flow rate ensures consistent and well-regulated sensing volume until the saturation capacity of the adsorbent pad (in the waste zone, Figure S2) is reached. In our case, a 0.10 mL saliva sample volume (Figure S10) to reach the sensing zone was sufficient for consistent sensing. As the signal started saturating after 5 minutes (Figure S11), a 0.10 mL sample and 5 minutes incubation time were optimized and considered for further experimentation. This information is highly recommended for the platform's practical usability. The saliva pH may vary in each individual based on their diet and health status [57], suggesting that the pH might affect sensor performance. The binding buffer's pH showed a maximum response at pH 7.4 (Figure S12), which is optimum for maximum sensor performance. These findings underscore the importance of the eMIP scaffold optimization for cortisol detection, a crucial aspect of our research.

### 2.4.2. Analytical Performances, reproducibility and selectivity Studies

To measure the salivary cortisol concentration promptly, the current change from amperometric (*i-t*) measurement is readout using a wireless potentiostat (Figure 5b). When a constant potential is applied, the current levels off for incubation beyond 60 sec, confirming the time needed to achieve saturated cortisol binding within the MIP cavities (Figure 5b). A calibration curve can be constructed between cortisol concentration and a settled value of the current in i-t-curves ($I_o$-I, µA). We further tested the sensing performance of saliva floss in 1X binding buffer (pH 7.4) with cortisol concentration ranging from 0.10 to 10000 pg mL$^{-1}$ for cort-eMIP (Figure 5c-d) and cort-eNIP (Figure 5e-f). The magnitude of the measured sensor current decreases as the amount of cortisol increases, indicating that the electron barrier is proportional to the amount of cortisol. This established a calibration curve showing a linear relationship between the current and cortisol concentration logarithm difference. For the cort-eMIP sensor, plotting the change in CA current relative to the baseline (without cortisol) against an increase in cortisol concentration yielded a linear response, Y (µA) = 0.8311 x $\log_{10}$ Cort(pg mL$^{-1}$) + 0.7972 in PBS with an excellent coefficient of correlation ($R^2$ = 0.9916, n=5) (Figure 5g, black curve). In contrast, eNIP-sensor showed a linear response with Y (µA) = 0.078 x $\log_{10}$ Cort(pg mL$^{-1}$) + 0.1016 ($R^2$ = 0.9898, n=5) in PBS (Figure 5g, red curve). The sensor performance was repeated for repeatedly measured for five different sensors, and the standard deviation was found to be less than 5.04% (n=5, for eMIP) and 5.69% (n=5, for eNIP). The results demonstrated that the saliva floss cortisol sensor has good selectivity and specificity for accurately detecting cortisol. The limit of detection (LOD) was calculated from the linear regression equation using the formula (LOD = 3.3 x σ/m), where m is the slope of the calibration curve, and σ is the standard deviation of the response from the regression equation in Figure 5g, resulting in a calculated LOD of 0.023 pg mL$^{-1}$ (in buffer medium). This LOD is more than sufficient for noninvasive cortisol detection in various biological fluids, including saliva, highlighting the practical applications of the saliva-sensing dental floss POC in healthcare diagnostics (see supplementary information ST1). The cort-eNIP saliva floss sensor showed minimal response to the same cortisol concentration, indicating the importance of

MIP cavities in the PPy film for the observed reaction (Figure 5g, red curve). This phenomenon can be explained by forming a denser PPy/PB film during e-NIP electropolymerization over cort-eMIP. It is important to highlight that the chronoamperometric measurements for the eNIP sensor are clearly different and do not overlap with those of the cort-eMIP sensor. The response of the cort-eMIP sensor was calculated and found to be approximately close to 10 % of the cort-eMIP sensor, highlighting the importance of enhanced sensitivity and high-precision performance of our developed platform. Furthermore, the relative reproducibility over five different devices at two different concentrations (10 and 1000 pg mL$^{-1}$) was also calculated to assess device-to-device variability. The high precision is further evident in Figure 5h, where CA measurements obtained from 5 different POCs at cort/PBS concentrations of 10.0 and 1000 pg mL$^{-1}$, with a sensor current at 60 seconds, show relative standard deviations (%RSD) of 3.78%, and 4.05 %, respectively.

Saliva encompasses a variety of small molecules, including hormones like cortisone, progesterone, testosterone, corticosterone, urea, glucose, lactate, etc., which share structural similarities with cortisol [10,58]. These analogous molecules can interfere with the sensor's operation by binding to the eMIP cavities, thus potentially modifying the sensor's signal. To ascertain the specificity of the cort-eMIP sensor, we performed CA current response measurements across a range of potential interferents, both structural and non-structural, at concentrations pertinent to physiological conditions. The relative change in the signal obtained with 1.0 pg mL$^{-1}$ cort was compared with the signal obtained against the common-interfering species at a higher concentration than cortisol. The relative change in the peak current obtained is presented in Figure 5i (with standard deviation, n=5). Cortisone, corticosterone, progesterone, and testosterone exhibited smaller responses than cortisol, possibly due to the presence of less hydroxyl functional moieties on the steroidal ring [59]. Additionally, cortisol contains a lactate moiety [37], which can allow the lactate molecule to have some binding and diffuse into the cavities. Glucose molecules, due to the presence of a number of hydroxy terminals, may interact with electron-rich sites (-NH group) and during charge transfer, affecting signal generation. However, no significant response was observed, evidenced by glucose and lactate showing a smaller response than urea; this could be due to the size difference and inappropriate binding sites. This comprehensive evaluation demonstrates our cortisol-sensing dental floss's high specificity and reliability, instilling confidence in its accuracy.

**Figure 5**

## 2.5. In-vitro testing

Before testing the saliva-sensing dental floss in humans, we conducted several experiments to confirm the ability to sample saliva and assess its ability to quantify cortisol (Figure 6a). Herein, we have used sulforhodamine (SRB, a dye) mixed with artificial saliva for visual confirmation of capillary microfluidic flow. As shown in Figure 6a, the developed saliva-sensing dental floss platform successfully collects the dyed artificial saliva sample with the microfluidic thread and transports the sample at 2 min. The microfluidic thread directs the sample toward the sensing zone, taking 5 min to reach, and to the waste zone, where it reaches in 7 min. This amount of time aligns with the time needed for incubation (see results in Figure S11). Subsequently, we assessed the sensing ability of the saliva-sensing dental floss using cortisol-spiked artificial saliva samples at concentrations that are representative of unstressed and stressed individuals, further demonstrating its practical application.

For this measurement, a lab-based dental model submerged in a cortisol-spiked artificial saliva sample was used (Figure 1b). The cortisol-spiked artificial saliva sample was sampled and allowed to flow toward the sensing zone (wait time of two minutes). After this incubation period, the sensing strip was inserted into the test zone, and the entire floss assembly was connected to the portable wireless readout device. A CA signal was recorded for varying cortisol concentrations (Figure 6b-c) in the above-mentioned settings. Each measurement was repeated five times to ensure the reliability of the signal. Figure 6d illustrates the calibration curve constructed using cortisol-spiked artificial saliva samples for both e-MIP and e-NIP devices. In artificial saliva, the sensitivity observed is slightly lower than that achieved in a cortisol-spiked buffer medium; however, it remains sufficient for deploying the developed platform as a saliva-POC diagnostic for cortisol monitoring. From Figure 6d, the cort-eMIP sensor showed a sensitivity of 160.08 nA/$\log_{10}$Cort (pg mL$^{-1}$), with a calculated LOD of 0.048 pg mL$^{-1}$ and $R^2$ = 0.9956 (n=5). In comparison, the cort-eNIP sensor showed a sensitivity of 15.26 nA/$\log_{10}$ Cort (pg mL$^{-1}$), with $R^2$ = 0.9948 (n=5). To the best of our knowledge, this is one of the lowest LODs achieved using a hand-handled saliva-based POC platform employing cort-eMIP/PPy/PB/PLEG-based electrodes for cortisol quantification over other reports as summarized (see supplementary table ST1). Additionally, we have also constructed a calibration curve in a narrow range (1, 2.5, 5, 7.5, and 10 ng/mL) targeting the cortisol levels in human saliva. The results are presented in the supplementary section (Figures S13 and S14) to distinguish between closely related cortisol concentrations. Our results indicate that the sensor demonstrates high precision and sensitivity, with a coefficient of variation (CV) ranging from 1.58% to 3.82% across different measurements, which falls within clinically acceptable limits. The sensor's response also shows a statistically significant distinction between these concentrations, validating its potential for accurate cortisol quantification in clinical applications with critical minor variations.

These findings hold significant implications for developing practical, real-world solutions in cortisol monitoring. It is important to note that the cortisol levels in saliva ranging from sub-picogram to nanogram per milliliter. However, our primary focus is to develop a platform capable of performing longitudinal measurements in stress monitoring, which requires monitoring subtle increases or decreases in cortisol levels as a result of therapy. The obtained lower LOD will allow for more accurate measurement to track changes in cortisol concentration with a change in behavioral intervention or lifestyle.

**Figure 6**

**2.6. Evaluation using spiked and real human saliva samples as part of a Stress Study**

The developed saliva-sensing dental floss platform was used to quantify cortisol concentration in a spiked human salivary sample at four different cortisol concentrations and validated against a cortisol ELISA kit. First, we recorded the blank signal (0.0 pg mL$^{-1}$ cort) to achieve a baseline. The blank showed a minimal current signal as output. Later, a 200 µL human saliva sample was spiked with known cortisol concentrations (10, 100, 1000, and 10000 pg mL$^{-1}$), mixed well, and divided into ELISA and sensor testing. Subsequently, the assays were carried out under the previously optimized conditions (as discussed above). Supplementary Table 2 (ST2) demonstrates that the recoveries for the cortisol concentrations were between 98.64% and 102.4 %, with their higher relative standard deviation of 1.55% for three consecutive measurements. This further confirms that our developed saliva dental floss can be used to determine cortisol in human saliva samples in a POC setting. It is evident from the results that the saliva-sensing dental floss developed in this work showed high precision and accuracy for cortisol quantification in human saliva samples.

From a clinical and pathological perspective, cortisol follows a diurnal pattern, and its levels in saliva fluctuate throughout the day, ranging from 1-12 ng mL$^{-1}$ in the morning to 0.1-3.0 ng mL$^{-}$

$^1$ in the evening [9,60]. These levels are significantly lower—about 100 times less than those found in blood, where cortisol levels are approximately 25 µg mL$^{-1}$ in the morning and 2 µg mL$^{-1}$ at midnight [9]. Therefore, to validate the performance and practical utility, the saliva-sensing dental floss platform performance was evaluated using real human saliva samples to assess cortisol levels, which were received as part of a Human Stress study (conducted at Tufts University as per the approved IRB protocol; see supplementary section S3). As mentioned above, from the clinical point of view, the salivary cortisol concentration ranges from sub-nanogram to nanogram per milliliter, which is equivalent to 1/10 of microgram per deciliter (µg/dl) in unit conversion. In ELISA analysis, the cortisol values are usually reported in the µg/dl; therefore, for comparison with the ELISA method, the cortisol values were calculated and reported in µg/dl. The procedure used for cortisol quantification is depicted in Figure 7a. Figures 7b and 7c summarize the results obtained from the quantification of human salivary cortisol levels as part of a longitudinal stress study. These samples were collected from human volunteers at various time points of a stress study, including baseline, immediately before stress induction, immediately after stress exposure, and 20 minutes post-stress, to investigate the change in stress response by evaluating cortisol levels. No flossing was performed by humans in this study, as this was out of the scope of the approved IRB protocol. Rather, the samples were manually applied to the dental floss to validate microfluidic flow and cortisol quantification with an eMIP sensor. Concurrently, the same samples were processed and subjected to analysis using a commercial cortisol-ELISA kit. It is important to note that the dilution factor was considered while calculating the cortisol's actual value in saliva. We calculated the regression between the cort-eMIP sensor as part of the dental floss and the ELISA kit. The results obtained, summarized in Figure 7b-c, show excellent agreement with less than an 8.0% standard deviation (ranging from 94.0% to 108.0%) for each of the testing methods.

These findings demonstrate that the designed sensing-floss cortisol sensing POC platform could reliably detect cortisol and exhibited strong to excellent consistency and agreement with the widely used ELISA method for detecting salivary cortisol (Figure 7d). There is a high correlation between the results obtained from the developed cort-sensing dental floss and ELISA sensing with a correlation factor of r = 0.9910 (with cortisol ELISA). The obtained comparative results further validate the accuracy of rapid cortisol detection using our proposed platform. The % relative error between cort-ELISA and saliva-sensing dental floss was found to be less than 5.0 %. This further confirms that our developed saliva-sensing dental floss, as POC, can be used to determine cortisol in human saliva samples. It is evident from the obtained results that the saliva-sensing dental floss developed in this work showed high precision and accuracy for cortisol quantification in actual human saliva samples, successfully validating its potential for routine point-of-care stress monitoring. In the future, we aim to measure cortisol levels in the general population, on wake up and before sleep, to establish a baseline and observe longitudinal trends; it will also include assessing the actual flossing performance in addition to sensing under a different IRB-approved study.

**Figure 7**

## 3. Conclusion and future outlook

The manuscripts report an innovative saliva-sensing dental floss intended to quantify cortisol levels in saliva and manage stress as part of one's daily routine. This saliva-sensing dental floss utilizes capillary action, as seen in hydrophilic threads, for saliva collection and transport. The portable device to hold the sensing dental floss is fabricated using a flexible biocompatible resin. The cortisol sensor component is fabricated on a flexible polyimide substrate using a combination of laser-engraved carbonization, which results in porous laser-engraved graphene (PLEG) electrodes and electrochemically polymerized molecularly imprinted polypyrrole (PPy/PB) to realize a cortisol-specific sensor (cort-eMIP/PPy/PB). Fabrication involves electrochemically depositing a PPy-PB scaffold onto the PLEG working electrode. By incorporating PB redox probes internally, the saliva-sensing dental floss could detect changes in the film's charge transfer process influenced by cortisol binding, resulting in a detectible redox signal. The porosity, high surface area, and high conductivity of PLEG results in superior sensor performance. It exhibited a linear response across a wide range of cortisol concentrations, ranging from 0.10 to 10000 pg mL$^{-1}$, covering levels found in healthy, unstressed, and stressed individuals. Notably, the saliva-sensing dental floss demonstrated the capability to detect cortisol levels as low as 0.023 pg mL$^{-1}$ in cortisol-spiked artificial saliva samples, which is among the best-performing cortisol sensors in the literature. This level of sensitivity allows monitoring of the smallest fluctuations in cortisol levels. The sensor is also highly selective due to the chemical selectivity of the cortisol-specific eMIP realized on the PLEG electrode. The sensor platform is versatile because it synthesizes and utilizes eMIPs that are selective to other targets. Furthermore, the MIPs can be engineered to detect different analytes of interest by templating the ideal conditions. There is also potential for multiplexed detection of several biomarkers of interest from saliva. Given its robust performance, this saliva-sensing dental flossing platform holds promise for widespread deployment as a point-of-care device for monitoring stress at the convenience of one's home. It is worth noting that the readout miniaturization and optimization were not part of the scope of the current work (first prototype). Moreover, future efforts may include further miniaturization and integration of the electronic functionality into the flossing device to improve compliance and ease of stress monitoring.

## 4. Experimental

### 4.1. Materials and Methods

Cortisol (human-certified standard) was purchased from Sigma Aldrich and used as received. Pyrrole (98%), potassium ferricyanide [$K_3Fe(CN)_6$], iron chloride hexahydrate ($FeCl_3 \cdot 6H_2O$), potassium chloride (KCl), hydrochloric acid (HCl, ACS reagent 37%), sodium dihydrogen phosphate ($NaH_2PO_4$), disodium hydrogen phosphate ($Na_2HPO_4$), and sodium chloride (anhydrous) were all purchased from Sigma Aldrich and used as received. Methanol (99%) was purchased from Thermo Scientific (USA). The cortisol competitive human ELISA kit was procured from ThermoFisher Scientific, USA. All the solutions were prepared in 0.20 µm membrane-filtered deionized water. BioMed Clear Resin was procured from FormLab, USA. Polyvinyl butyric acid (PVB, Butavar (B-98) was procured from Benton Chemical, USA, and used as received. The silver/silver chloride (Ag/AgCl) conductive ink for coating connection pads was procured from Kayaku Advanced Materials (USA). The polyimide sheet (PI) (thickness 12.5 µm, size 1.0 cm x 1.0 cm) was procured from Grainger, USA, and used as received. Testosterone, corticosterone, glucose, interleukin-6 (IL6), and lactate were purchased from Sigma Aldrich (USA) for interferent studies.

### 4.2. Instrumentation

The surface morphology of the fabricated sensor at different stages was characterized using Nicolet 6700 FTIR attached with Smart iTX (Thermo Scientific) in ATR mode. Scanning electron microscope (Axia), Thermo Scientific (USA) used to generate SEM images. The portable potentiostat (Electrochemical Workstation) was purchased from Zensor (Thailand). A laser engraving and cutting system, equipped with a $CO_2$ laser of wavelength $\lambda = 10.6$ µm, was employed to fabricate PLEG under various conditions. These conditions involved the variation of laser power (P), scanning speed (v), repeated scanning passes (n), off-focus value (If), image resolution, and gas flow. Raman spectrum was recorded on a DXR Raman microscope (Thermo Scientific) at 532 nm. A microplate reader (Thermo Scientific) was used for ELISA measurements.

### 4.3. Sensing-floss design and fabrication

#### 4.3.1. Design and Fabrication of Sensing dental floss

The step-by-step fabrication and assembly of the saliva-sensing dental floss design are shown in Figure 1(c-k) and Figure S1. Initially, the 3D design for the primary floss (base layer) and the cover (top layer) was created using AutoCAD Fusion 360 software and saved as a mesh as an STL file. Subsequently, this STL file was exported to the PreForm software and transferred to the FormLabs 3B 3D printer, where the main floss structure and cover are printed using BioMed Clear Resin (Figure 1c). The BioMed Clear was chosen for its biocompatibility and has been used extensively in dentistry. Upon completion of 3D printing, the structures were detached from the built platform and immersed in isopropyl alcohol (IPA) for 30.0 minutes and then cured at 60.0 °C for 60.0 minutes. Concurrently, a surgical adsorbent pad was cut and affixed to the main floss structure using a binder to serve as a waste saliva collection zone (Figure S1b). The pad also ensures continuous unidirectional flow of saliva samples from the floss toward the waste zone. For the microfluidic region of the platform, a cotton thread is pre-treated with 0.05 M HCl for 5.0 minutes, dried at 50.0 °C for 20.0 minutes, washed with ethanol for 10.0 minutes, and dried at 40.0 °C for 20.0 minutes before use (Figure 1d). Plasma treatment for 30.0 minutes in two intervals of 15.0 minutes each rendered the thread hydrophilic. A segment of the thread was cut, and a single fiber was extracted from the bundle. This fiber was then inserted into the holes of the main floss design, and using a glue gun; it was affixed so that the glue only made contact with the thread in contact with the structure (Figure S1c). To avoid interference with capillary action, care was taken not to let the glue touch the thread where capillary action was essential. Subsequently, the microfluidic thread was cut and placed to fit into the grooves of the structure, touching the primary floss fiber to enable capillary action with minimal hindrance (Figure 1f and Figure S1d). Following this, the second half of the adsorbent pad was attached, sandwiching the microfluidic thread. The entire setup was then subjected to a plasma cleaner for an additional 15.0 minutes. Finally, the 3D-printed cover was affixed to the main floss structure. In parallel, porous laser-engraved graphene (PLEG) electrodes (Figure 1e) were fabricated through the laser engraving method (as discussed in the next section). A 3D assembly of sensing dental floss is depicted in Figure 1f, and an assembled cort-floss sensing device is shown in Figure 1g.

### 4.3.2. Fabrication of PLEG electrodes

Laser-induced carbonization of thin, flexible polyimide (PI) surface is used to realize the PLEG electrodes (Figure 1e) for working (WE), counter (CE), and reference (RE) electrodes (Figure S4). The electrode design used in this work is a conventional 3- 3-electrode system made up of working, counter, and reference electrodes. The diameter of the working electrode is 3 mm. First, a cleaned PI sheet was secured onto a laser bed using a transparent adhesive tape and exposed to the laser (laser power- 30%, speed- 5.2%). Nitrogen ($N_2$) gas assists with the flow at 50% and is applied at the engraving site. The PLEG electrodes had an average resistance of 38 ± 3 Ω/cm (n=10). The resulting PLEG electrodes show high surface area and excellent electrical conductivity and are amenable to chemical surface modification to realize sensors [44,45]. Subsequently, the RE and the external contact pads underwent Ag/AgCl conductive paste coating. The RE was further stabilized by drop-casting polyvinyl butyrate (PVB, containing sodium chloride dissolved in methanol) and dried (24 hr.). Before use, the electrode underwent a surface cleaning and activation by performing a voltage sweep between 1.0 and -1.50 V at a scan rate of 100 mV/ sec in 0.50 M $H_2SO_4$ (containing

0.10 M KCl as supporting electrolyte) until a clean cyclic voltammetry (CV) curve is not obtained (Figure S5).

### 4.3.3. Fabrication of eMIP-based cort-sensor

The cort-eMIP sensor fabrication starts with electropolymerization of a pyrrole-based polymeric cocktail containing template (cortisol) in phosphate-buffered saline (PBS, pH 7.4) by sweeping cyclic potential from -0.30 to 0.80 V at a scan rate of 50 mV/s for 20 repetitive cycles. The polymerization cocktail solution comprised pyrrole (20 mM), cortisol (5 mM), $FeCl_3.6H_2O$ (5.0 mM), $K_3[Fe(CN)_6]$ (5.0 mM), and HCl (100 mM) until not optimized. The chemical structure of cocktail components is depicted in Figure 2a. Figure 2b represents an image of a fabricated PLEG electrode. During voltammetric oxidation of the polymeric cocktail (pyrrole, prussian blue components, and template (cortisol)), the hydrogen bonding interaction between electronegative oxygen-containing functional groups of cortisol and the electropositive hydrogen present nitrogen of the pyrrole molecule helps encapsulate cortisol into the PPy-PB network during electropolymerization (Figure 2c-d). Post electropolymerization, the cort-eMIP/PPy/PB/PLEG electrode was washed three times with PBS to remove unreacted monomer or template molecules. Later, cort-eMIP/PPy/PB/PLEG electrodes were over-oxidized in PBS from -0.40 to 0.60 V for 20 cycles to extract cortisol molecules. Overoxidation of the eMIP scaffold led to the formation of cort-binding (cort-memory) sites within the PPy/PB matrix. Similarly, the non-imprinted polymer (cort-eNIP/PPy/PB/PLEG) was fabricated without the template (cortisol) in the polymeric mixture. The step-by-step mechanism of cort-eMIP/PPy/PB/PLEG sensor fabrication is depicted in Figure 2c.

### 4.3.4. Solution preparation

The performance and applicability of our developed saliva-sensing dental floss for cortisol were performed in the binding buffer and artificial saliva samples. The preparation of binding buffer and artificial saliva solutions are described in detail in the supplementary information (section S1).

### 4.3.5. Matrix (Human Saliva) collection and sample preparation for cortisol detection

The human saliva samples used in this work were collected as per the approved protocol from the Institutional Review Board (IRB), Tufts University (Protocol No. 1706030). The detailed description is provided in the supplementary section (S2).

### 4.3.6. Sensor readout (Chronoamperometric measurement)

To emphasize the interfacial properties of the fabricated cort-eMIP sensor during each fabrication step, cyclic voltammetry (CV) and electrochemical impedance spectroscopy (EIS) were performed in 1.0 mM ferricyanide/ferrocyanide redox medium prepared in 0.10 M PBS (pH 7.4, 0.10 M KCl) buffer. Leveraging the advantage of embedded internal redox species (PB), sensor sensitivity was validated using chronoamperometry (CA) to detect a relative change in the PB oxidation current resulting from the occupation of memory sites by cortisol. After flossing, which represents the time needed for the saliva sample to travel through the thread-microfluidic channel and reach the horizontally supported 3-electrode sensing surface, the sensor electrode was inserted into the sensing port and allowed to equilibrate for an additional 3 minutes. Subsequently, a CA signal was applied under the following parameters: initial potential 0 V, high potential 0.10 V for a pulse width of 60 sec, with a sample interval of 0.10 seconds. The change in oxidation current was measured after the signal stabilization.

**Conflicts of interest**

The author declares no conflict of interest.

**Author contributions**

**Atul Sharma:** Conceptualization, Methodology, Investigation, Experimentation, Data curation, Validation, Formal Analysis, Writing-Original draft preparation. **Nafize Ishtiaque Hossain**: Fabrication support, Data Curation, Visualization, Software. **Ayanna Thomas:** Resources, Writing-Reviewing, and Editing. **Sameer Sonkusale:** Resources, Supervision, Project Administration, Funding Acquisition, Writing-Reviewing, and Editing.

**Acknowledgments**

This research received funding from the National Science Foundation (NSF) under Grant 1931978. Additionally, it was partially supported by the TTW program at the Uniformed Services University of Health Sciences (USUHS) with award HU0001-20-2-0014 and the DoD Peer Review Medical Research Program with award number W81XWH-21-2-0012. The collection of human samples adhered to the protocol approved by the Institutional Review Board (IRB) at Tufts University (1706030).

**Figure and Tables**

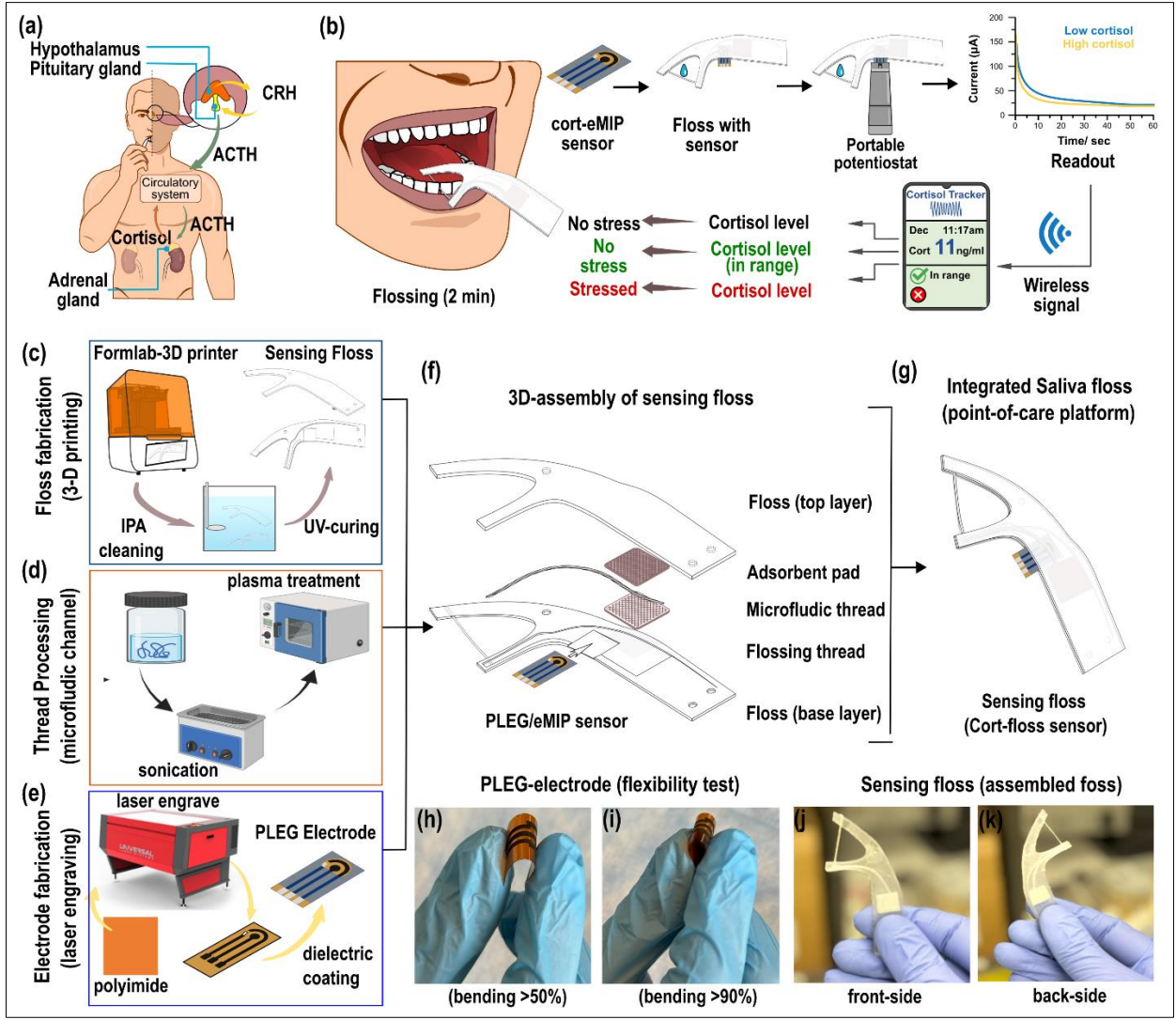

**Figure 1:** (a) Role of cortisol in stress response; (b) Proposed saliva sensing dental floss use case with the embedded cort-eMIP sensor. The floss is connected to a portable potentiostat for readout and signal transmission to a mobile device; Step-by-step process flow for fabrication of sensing dental floss device (c) fabrication of floss package (3D printing); (d) preparation of flossing and microfluidic thread; (e) fabrication of PLEG electrode through laser engraving; (f) 3D- stack assembly of floss device; (g) assembled saliva sensing dental floss device; micrograph of a fabricated electrode at the bending of (h) 50 % and (i) > 90% angles; images of 3D fabricated sensing dental floss (j) front side; (k) backside.

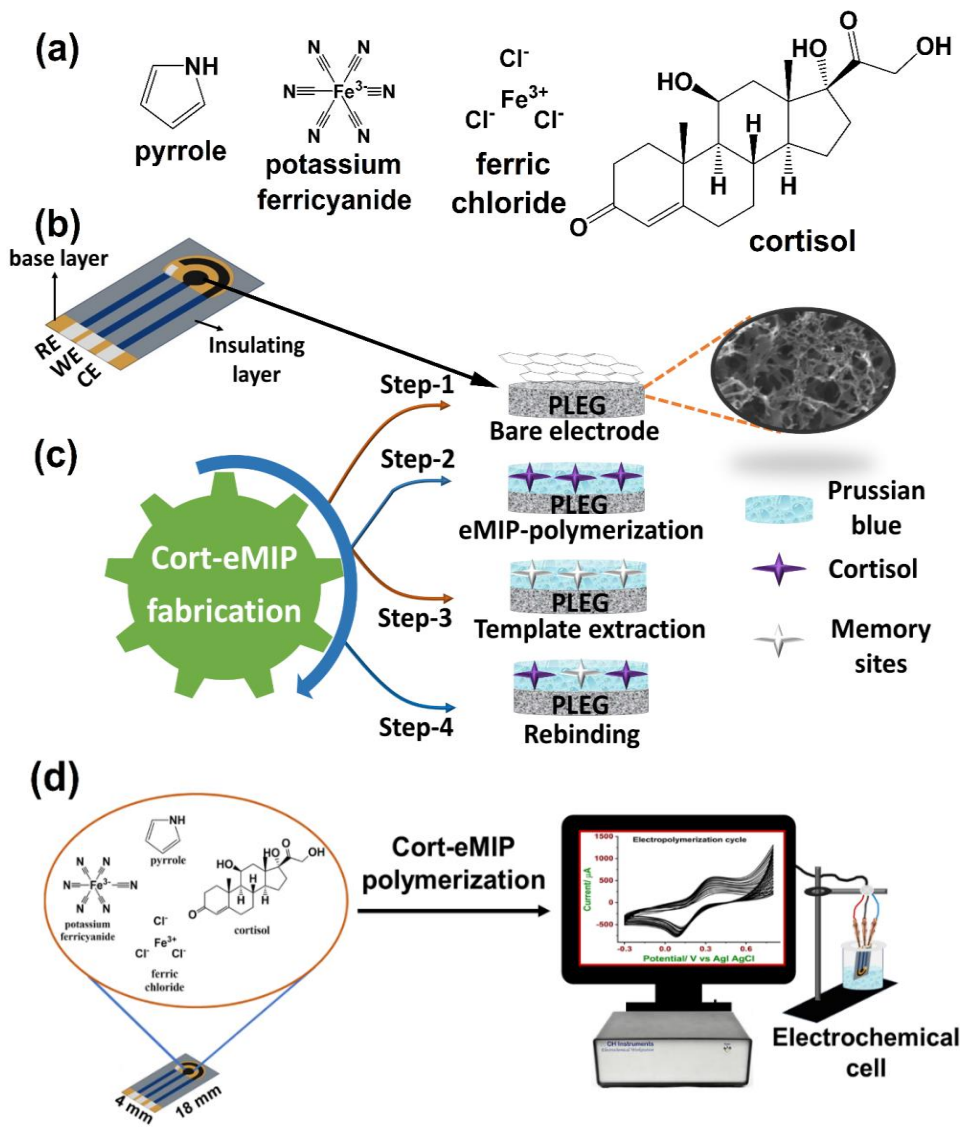

**Figure 2:** (a) Components of polymeric cocktail used; (b) image of 3-electrode; (c) step-by-step fabrication of the cort-eMIP sensor; (d) electropolymerization for preparation of cort-eMIP sensor

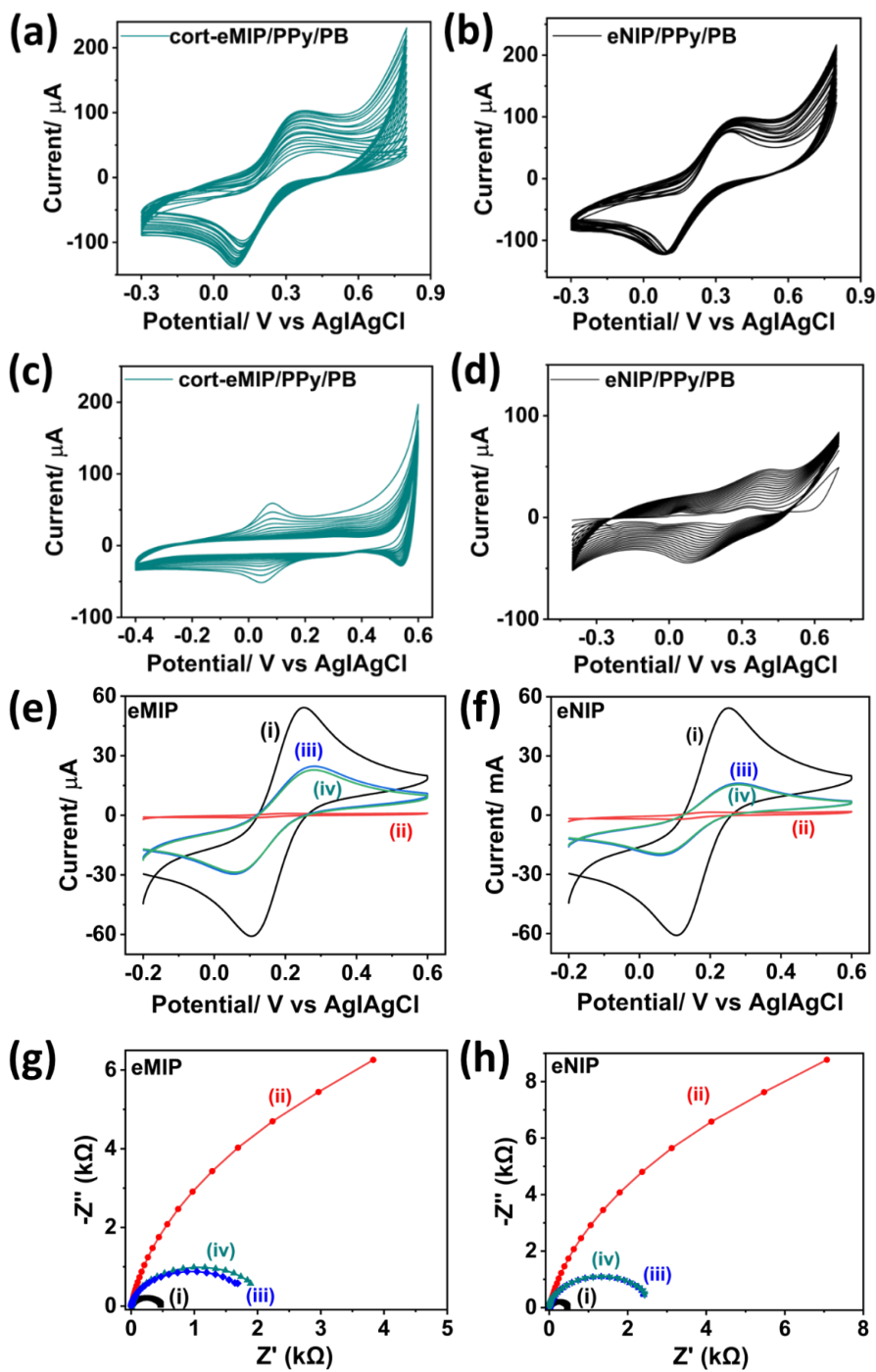

**Figure 3:** Electropolymerization of polymeric cocktail containing (pyrrole, ferric chloride hexahydrate, potassium ferricyanide) for the fabrication of (a) cort-eMIP/PPy/PB sensor (with cortisol); (b) cort-eNIP/PPy/PB sensor (without cortisol) with sweeping potential over -0.30 to +0.80 V (20 cycles), at 50 mV/s in phosphate-buffered saline (PBS) medium (0.10 M, pH- 7.4). Cyclic voltammograms of template extraction (overoxidation process) of (c) cort-eMIP/PPy/PB sensor and (d) cort-eNIP/PPy/PB sensor in PBS (0.10 M, pH 7.4). CV and EIS characterization of (e and g) cort-eMIP/PPy/PB sensor; (f and h) cort-eNIP/PPy/PB sensor at different steps of fabrication in the presence of 1.0 mM ferrocyanide/ ferricyanide couple prepared in PBS (0.10 M,

pH 7.4) for different steps, i.e. (i) bare PLEG, (ii) electropolymerized, (iii) after template removal, and (iv) after rebinding of 1.0 pg mL$^{-1}$ cortisol.

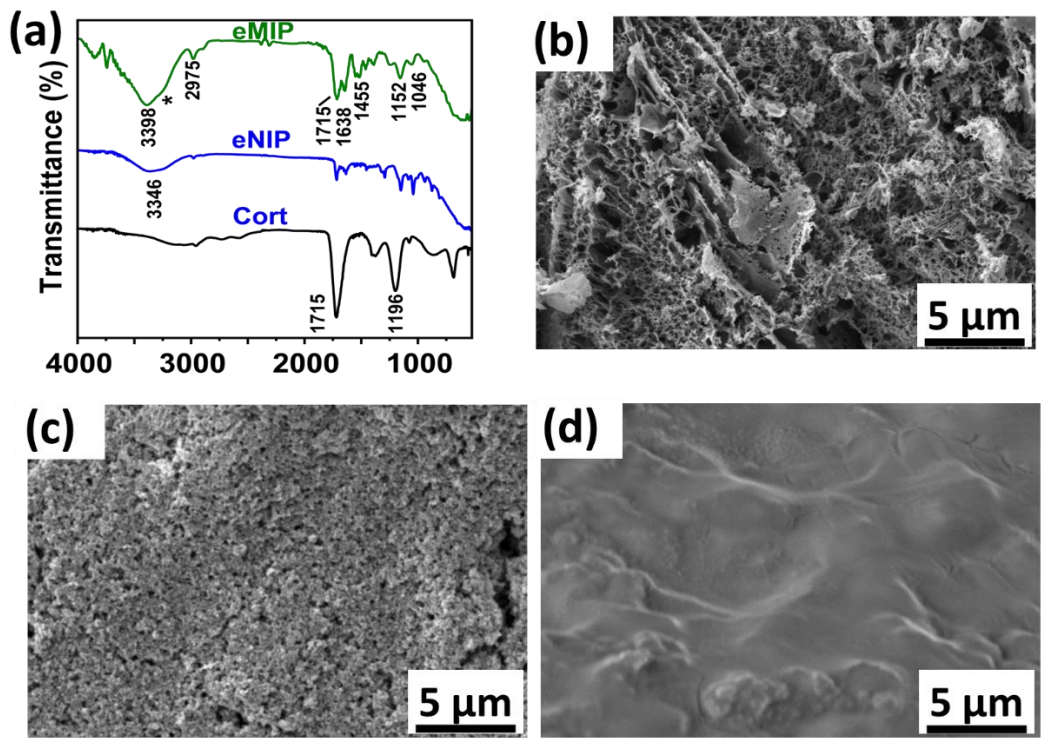

**Figure 4: (a)** FTIR spectrum of PLEG electrode (black curve), cort-eNIP/PPy/PB/PLEG electrode (blue curve), cort-eMIP/PPy/PB/PLEG electrode (green curve) recorded at a scan rate of 32 cm$^{-1}$ in ATR mode; SEM micrograph of **(b)** PLEG electrode; **(c)** cort-eMIP/PPy/PB/PLEG electrode; **(d)** cort-eNIP/PPy/PB/PLEG electrode.

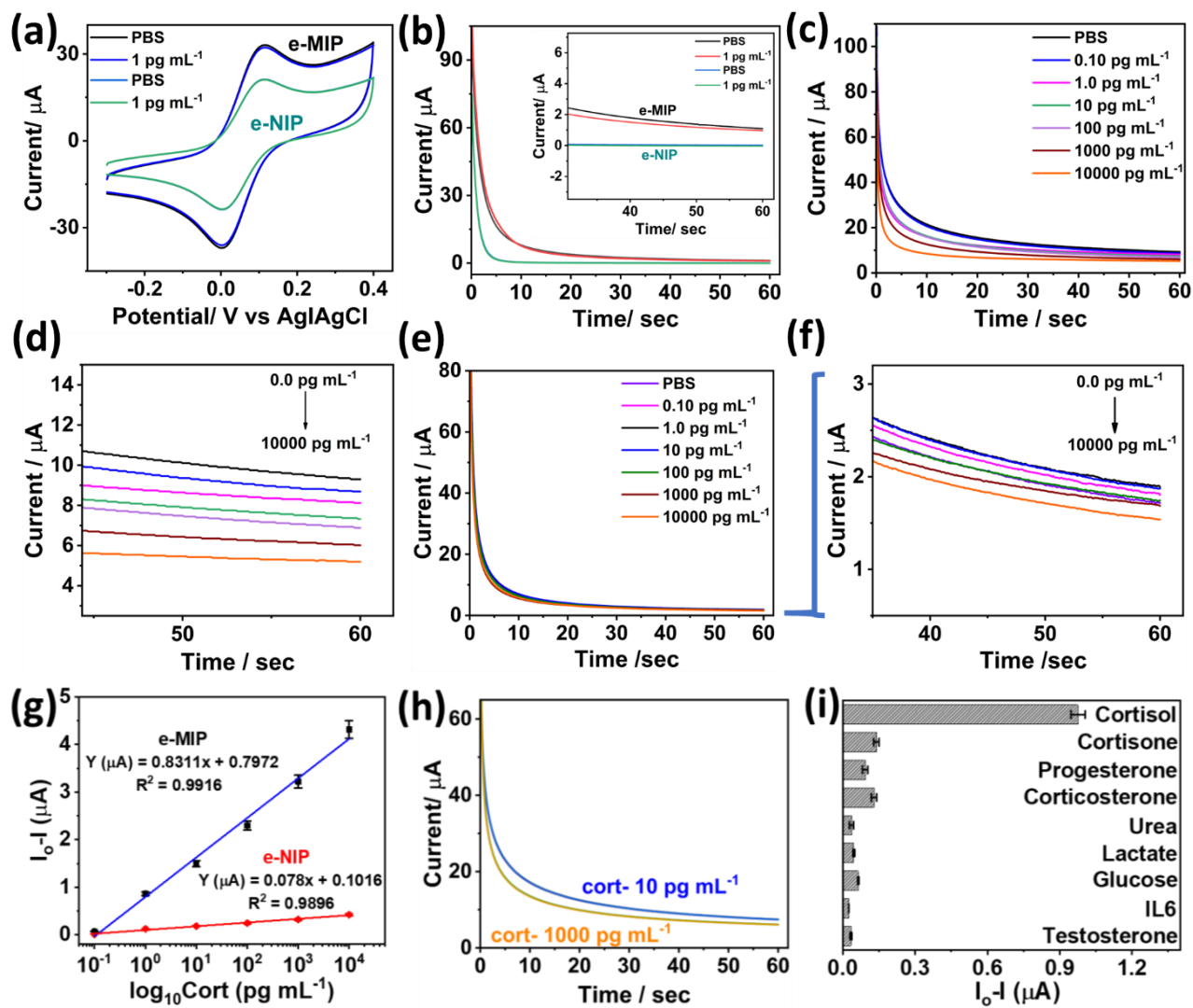

**Figure 5:** Characterization of cort-eMIP and cort-eNIP sensor performance. (a) Cyclic voltammogram (CV) with an oxidation peak at around 0.10 V with a scan rate of 0.10 V/s; and (b) Chronoamperogram (CA) with oxidation current magnitude at 0.10 V bias for cort-eMIP and cort-eNIP with and without 1.0 pg mL$^{-1}$ cort concentration. (c) Chronoamperogram (CA) of the cort-eMIP and cort-eNIP sensor with increasing cortisol concentration (0.10 to 10000 pg mL$^{-1}$) showing a decrease in oxidation current magnitude at 0.10 V bias; (d) magnifying CA curve of eMIP sensor; (e) CA curves cort-eNIP sensor with increasing cortisol concentration (0.10 to 10000 pg mL$^{-1}$) showing a decrease in oxidation current magnitude at 0.10 V bias; (f) magnifying CA curve of eNIP sensor; (g) Calibration curve of cort-eMIP and cort-eNIP sensor in PBS (0.10 M, pH 7.4) with cortisol ranges from 0.10 to 10000 pg mL$^{-1}$; (h) reproducibility performance of the sensor at 10 and 1000 pg mL$^{-1}$ (n=5); (i) interference and selectivity studies with structural (cortisone, progesterone, testosterone, and corticosterone) and non-structural (glucose, lactate, IL6, urea) analogs. Each plot has an error bar depicting the replicate measurements of n = 5 sensors.

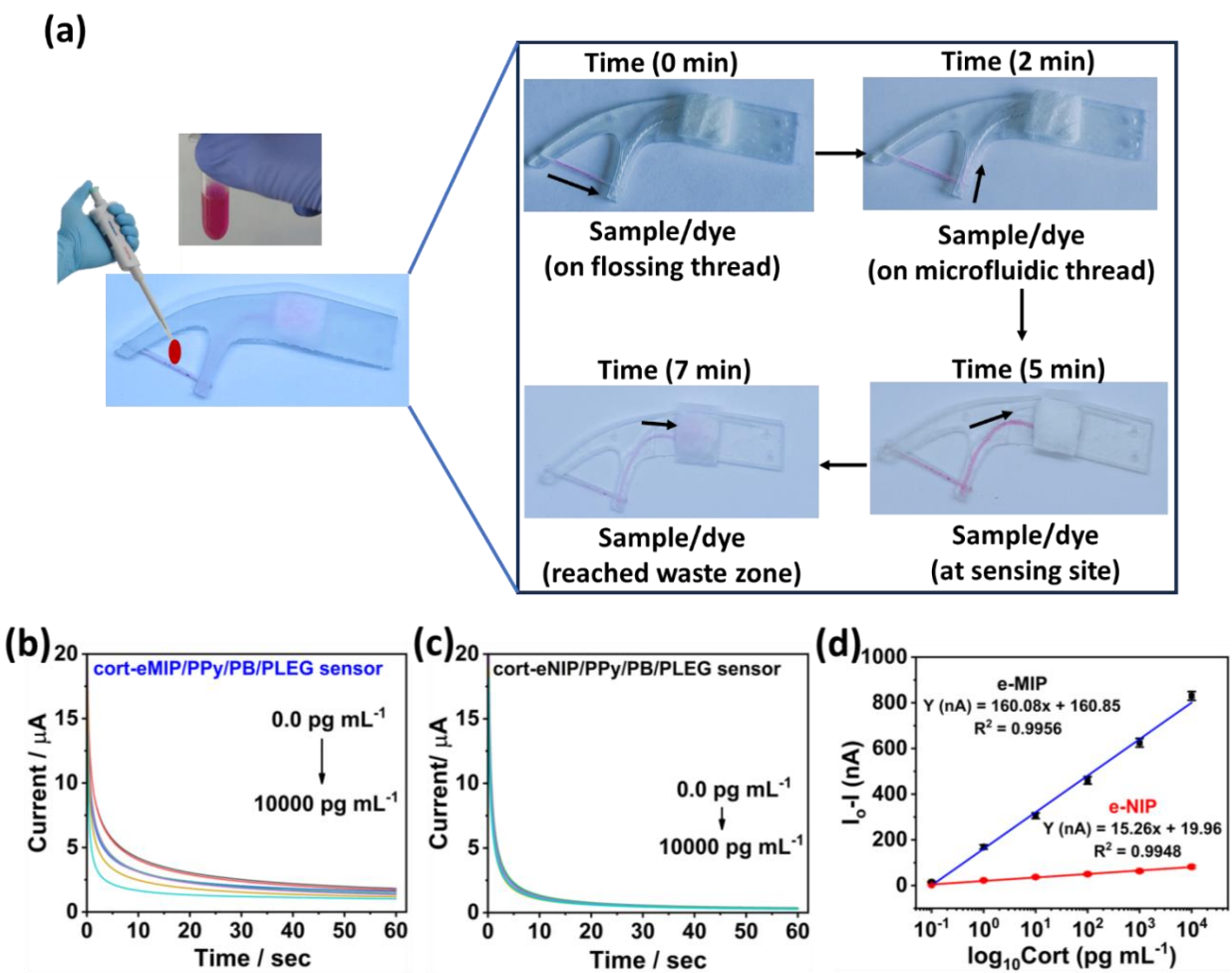

**Figure 6:** (a) Demonstration of sampling (saliva, 0.10 mL) and flow of saliva from flossing site to waste zone using sulforhodamine B (SRB, a dye) for visual confirmation; Chronoamperogram (CA) for (b) cort-eMIP and (c) cort-eNIP with increasing cortisol concentration (0.10 to 10000 pg mL$^{-1}$) showing a decrease in oxidation current magnitude at 0.10 V bias in cortisol-spiked artificial saliva samples; (d) Calibration curve of cort-eMIP and cort-eNIP sensor in cortisol-spiked artificial saliva samples with cortisol ranges from 0.10 to 10000 pg mL$^{-1}$. Each plot has an error bar depicting the replicate measurements of n = 5 sensors.

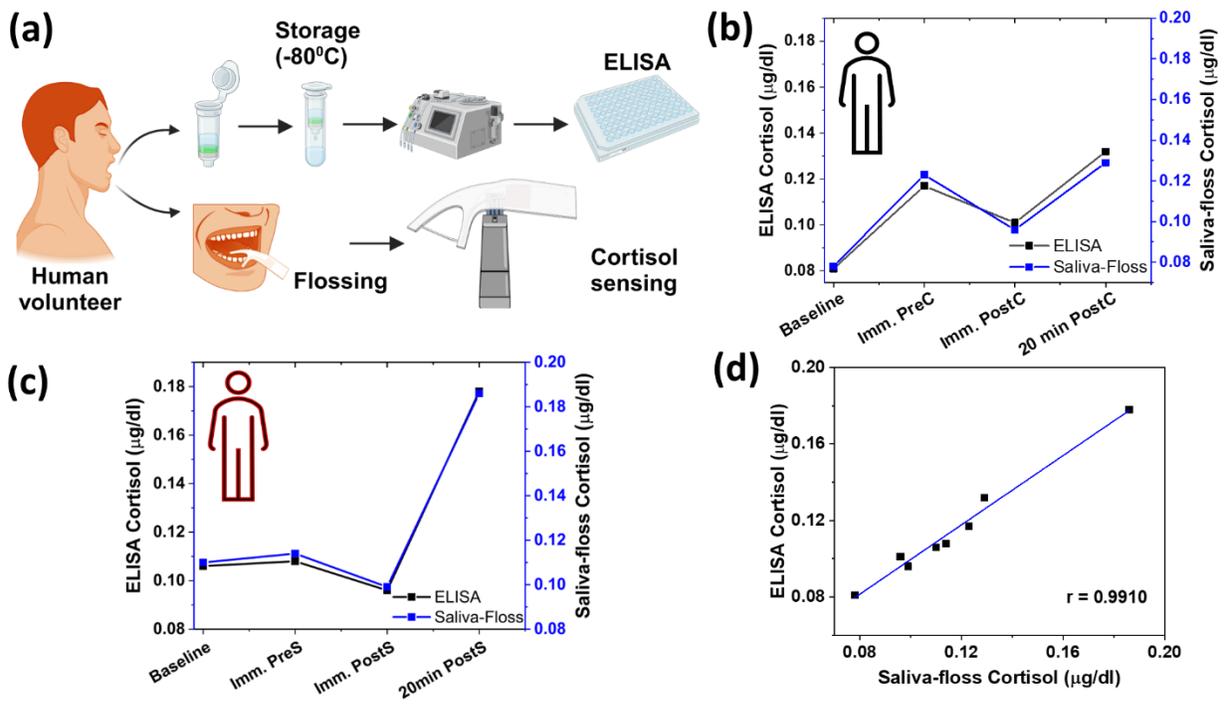

**Figure 7:** (a) Process flow of human saliva collection and corresponding screening methods; quantification of salivary cortisol using ELISA and saliva sensing dental floss in (b) in control (c) in stress-induced student volunteer; (d) establish a correlation between ELISA and saliva sensing dental floss for salivary cortisol quantification. Note: Imm.: immediately; Imm. PreC: immediately pre-control; Imm. PostC: immediately post-control; Imm. PreS: immediately pre-stress; Imm. PostS: immediately post-stress.

# Saliva-Sensing Dental Floss: A Novel Tool for Assessing Daily Stress via On-Demand Salivary Cortisol Measurement Integrating Molecularly Imprinted Polymer and Thread Microfluidics

Atul Sharma [a,b], Nafize Ishtiaque Hossain [a,b], Ayanna Thomas [c], Sameer Sonkusale [a,b*]

[a] Nano Lab, Advanced Technology Laboratory, Tufts University, Medford, MA, USA.
[b] Department of Electrical and Computer Engineering, Tufts University, Medford, MA, USA
[c] Department of Psychology, Tufts University, Medford, MA 02155, USA
*Corresponding author email: sameer@ece.tufts.edu, sameer.sonkusale@tufts.edu

# Table of Contents

**Supplementary Figure S1**

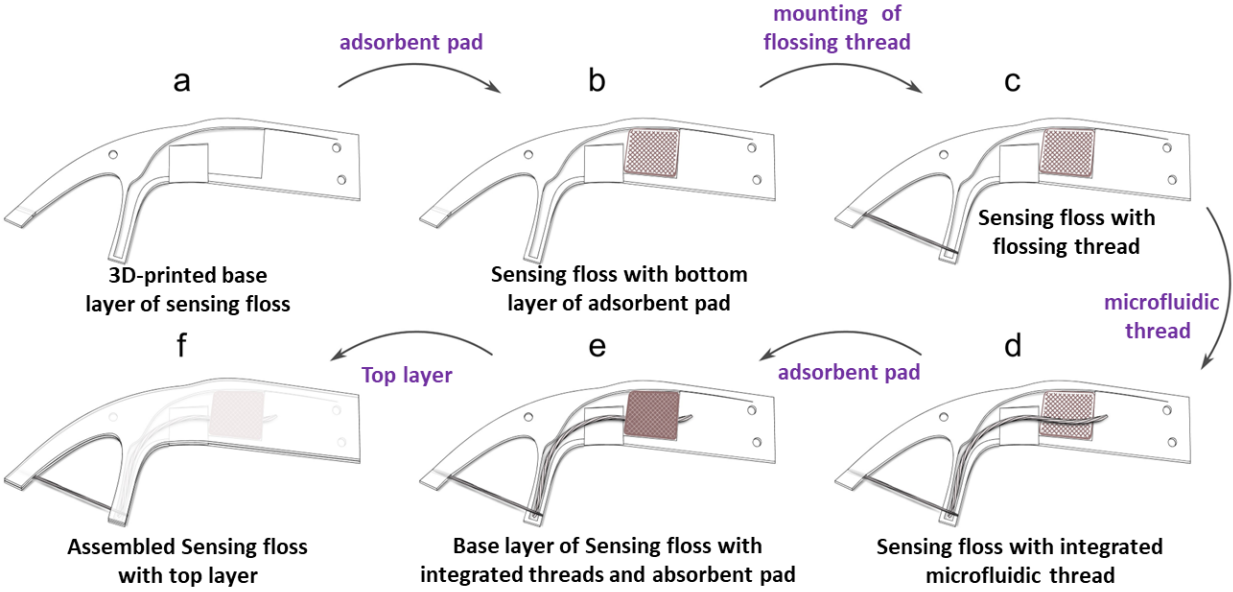

**Figure S1**: Step-by-step assembly fabrication process of saliva-floss

**Supplementary Figure S2**

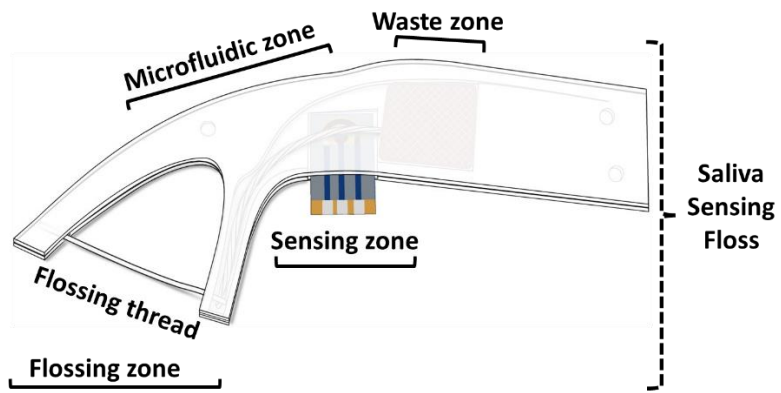

**Figure S2**: Saliva-floss showing different zones (Flossing, microfluidic, sensing, and waste zone)

**Supplementary Figure S3**

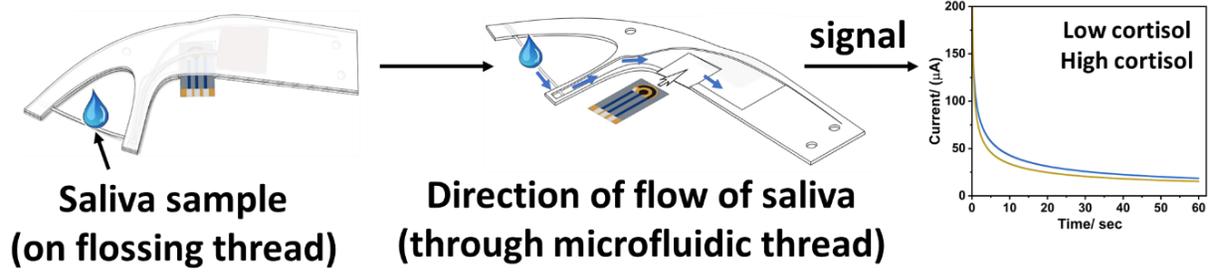

**Figure S3:** Flow of saliva sample towards sensing zone (or sensing strip) and corresponding signal generation concerning cortisol level

**Supplementary Figure S4**

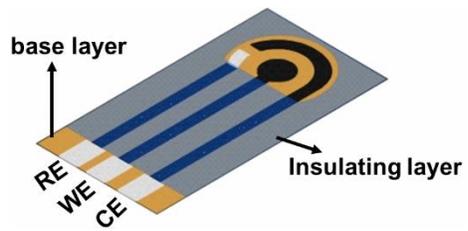

**Figure S4**: Porous laser engrave graphene (PLEG) 3-electrode (WE: working electrode; RE: reference electrode; CE: counter electrode). The diameter of the working electrode is 3 mm.

**Supplementary Figure S5**

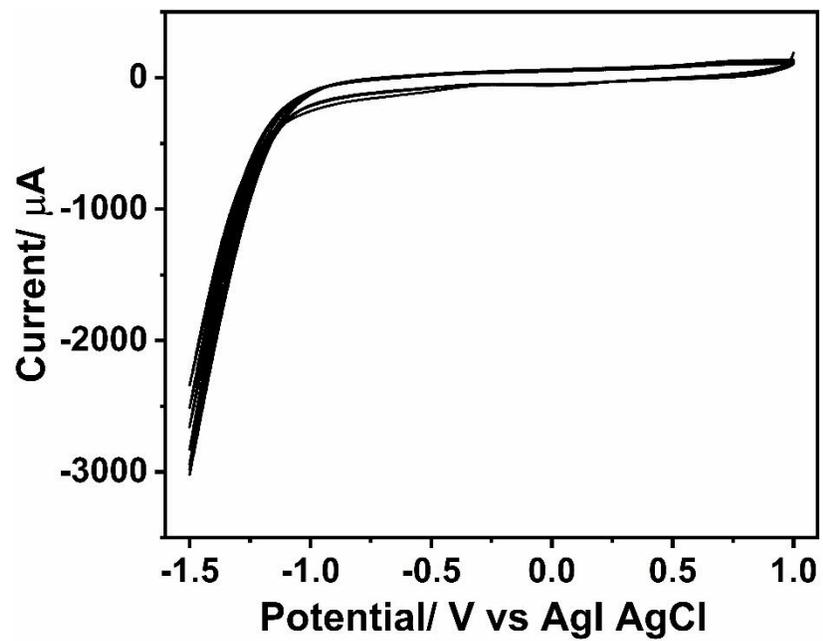

**Figure S5**: Cyclic voltammogram of PLEG electrode by sweeping the voltage from 1.0 V to -1.5 V at 100 mV/sec in 0.50 M sulphuric acid containing 0.10 M potassium chloride

**Supplementary section (S1): Preparation of solution**

1. **Phosphate binding buffer (0.10 M, pH 7.4)** was prepared by mixing $Na_2HPO_4$ and $NaH_2PO_4$ with deionized (DI) water and then adjusting the pH with NaOH (0.10 M) and/or HCl (0.10 M).

2. **Artificial saliva samples were prepared** by mixing $Na_2HPO_4$ (2.40 mM), $KH_2PO_4$ (2.50 mM), $KHCO_3$ (15.0 mM), NaCl (10.0 mM), $MgCl_2$ (1.50 mM), $CaCl_2$ (1.50 mM) and critic acid (0.15 mM), subsequently adjusting to pH 7.4 with NaOH (0.10 M) and/or HCl (0.10 M).

**Supplementary section (S2): Human saliva collection and storage**

Human saliva samples were collected and obtained as per the approved protocol from Tufts University's Institutional Review Board (IRB) under Protocol No. 1706030 for another NSF-funded project. Human saliva samples of student volunteers as a part of the stress study were collected at Tufts University using validated polypropylene vials, specifically SalivaBio 2ml cryovials (Salimetrics Item No 5004.01), through the passive drooling method. The collection occurred at four distinct time points (baseline, immediate pre-stress, immediate post-stress, and 20 min post-stress) to investigate the stress response. After collection, the vials were securely sealed to withstand temperatures as low as -80 °C and featured external threading to facilitate the use of the Saliva Collection Aid (SCA- Salimetrics Items No. 5016.02) for guiding drool directly into the cryovial. Participants inserted a small cotton tube into their mouths and sucked on it for one minute, then placed the cotton piece into a tube for later testing. Subsequently, the tubes were transferred a conical tube storage box and stored in a laboratory refrigerator set at -80 ºC for preservation. All salivary cortisol underwent testing using the enzyme-linked immunosorbent assay (ELISA), a standard testing method, within three months of collection. Samples were shipped frozen to Salimetrics for ELISA analysis via FedEx. The saliva collected from the bottom of the V-tube underwent further testing with the Cort-sensor and a commercial cortisol ELISA-Kit.

The saliva sample obtained from the lower part of the salivary vessel was subjected to centrifugation at 2500x rpm for 20 minutes. The resulting supernatant was collected, mixed with an equal volume of PBS (at a 1:1 ratio, v/v), and thoroughly homogenized before usage. A 100 µL aliquot of the saliva sample was dropped to the flossing thread surface for cortisol sensing. At the same time, the second portion was utilized for cortisol ELISA-kit testing to validate the results. The dilution factor was considered when determining the actual concentration of cort in the saliva sample. After the testing, the sensors and materials that had come into contact with the saliva samples were handled in compliance with regulatory guidelines to mitigate potential biohazard risks.

**Supplementary Figure S6**

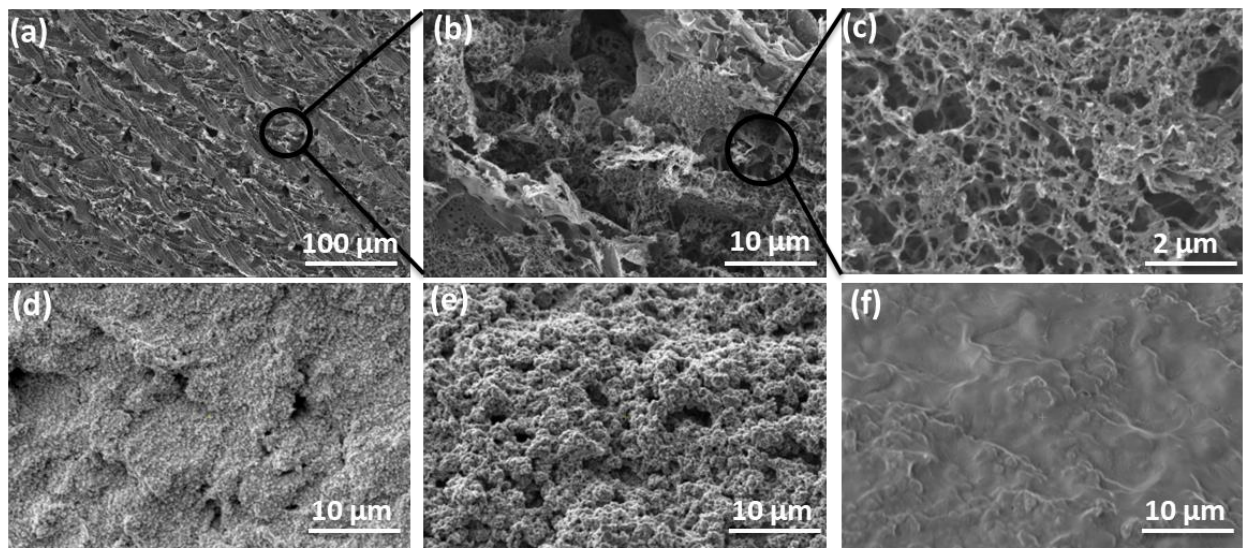

**Figure S6:** SEM images of (a) cleaned PLEG-electrode (at 100 µm); (b) zoom-in image of PLEG-electrode (at 10 µm); (c) PLEG-electrode (at 2 µm); (d) cort-eMIP electrode before extraction (at 10 µm); (e) after extraction of electropolymerized Cort-eMIP on PLEG electrode (at 10 µm); (f) after extraction of electropolymerized Cort-eNIP electrode (at 10 µm).

**Supplementary section (S3): Raman analysis**

The Raman spectrum of the porous laser-engraved graphene (PLEG) electrode and polyimide (PI) surface were compared to study and confirm the graphene production. The Raman spectrum of the PLEG electrode exhibited three distinguished peaks (D, G, and 2D peaks) (Fig. 3a, blue curve). In contrast, the non-engraved PI surface did not show any prominent band (Fig. 3a, black curve), which is usually present within 1200-1300 cm$^{-1}$. Small spikes close to the baseline between 1800 to 2500 cm$^{-1}$ are due to the background signal. This observation is consistent with findings from Lin et al. and others in the literature [1,2]. The ratio of IG/ID usually characterizes the degree of disorderliness and defects. After a thorough analysis, it is essential to highlight the increase in the variation of the $I_G/I_D$ ratio for the PLEG electrode ($I_G/I_D$ = 0.80), cort-eMIP/PLEG electrode ($I_G/I_D$ = 0.833), and cort-eNIP/PLEG electrode ($I_G/I_D$ = 0.805), which suggests an increase in the number of defects in the surface post eMIP formation. These defects imply that there has been a modification in the working electrode with the polymerization process. The polymer appears to have fewer defects similar to those of the PLEG electrode. This seems surprising considering that PLEG is a highly porous structure with defects, whereas eNIP does not show more defects due to the formation of almost negligible cortisol binding sites. If we compare the ratio of the NIP with the MIP, this eMIP has more defects, which is to be expected since the eMIP matrix contains the cortisol that can cause defects in the polymer post template extraction. Additionally, the ID/IG ratio of the MIP sample further decreased after the cortisol reading compared to other samples associated with cortisol extraction due to cavities in the polymer.

**Supplementary Figure S7**

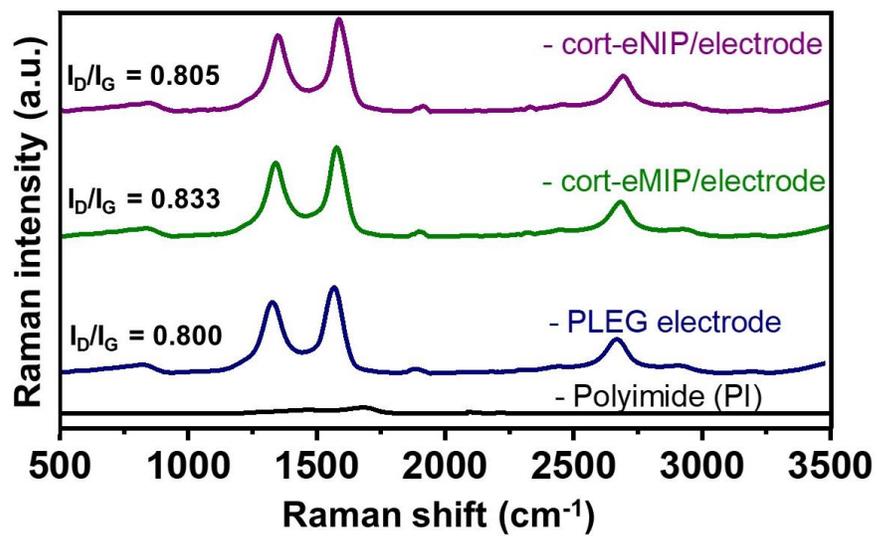

**Supplementary Figure 7 (S7):** Raman spectrum of PLEG-electrode (black curve); polyimide (PI, blue curve); cort-eMIP electrode (curve green), and cort-eNIP electrode (violet curve)

**Supplementary Figure S8**

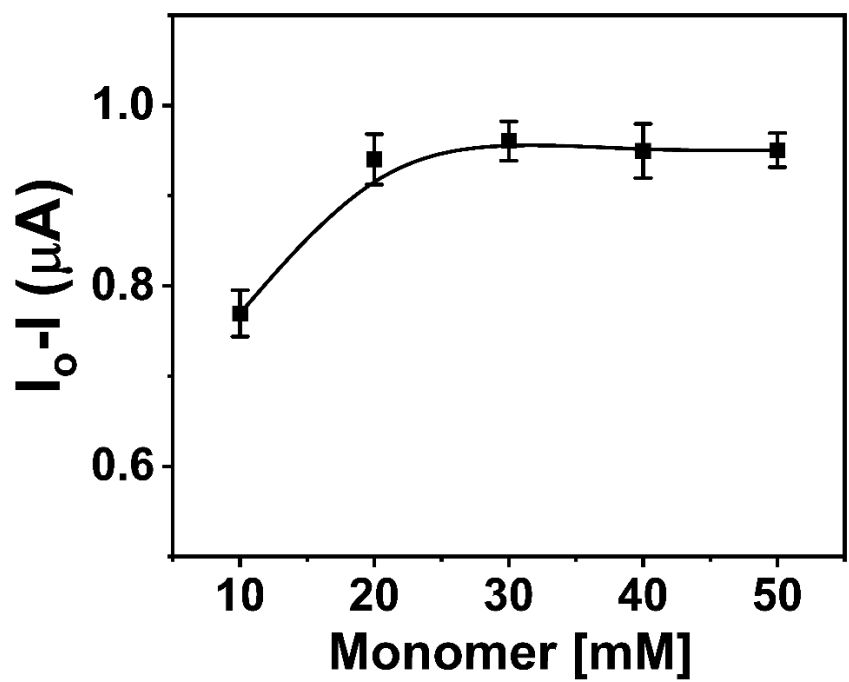

**Figure- S8:** Effect of functional monomer (pyrrole) concentration with an increase in concentration from 10 mM to 50 mM in the presence of cortisol (1.0 pg mL$^{-1}$) in binding buffer (0.10 M, pH, 7.4).

**Supplementary Figure S9**

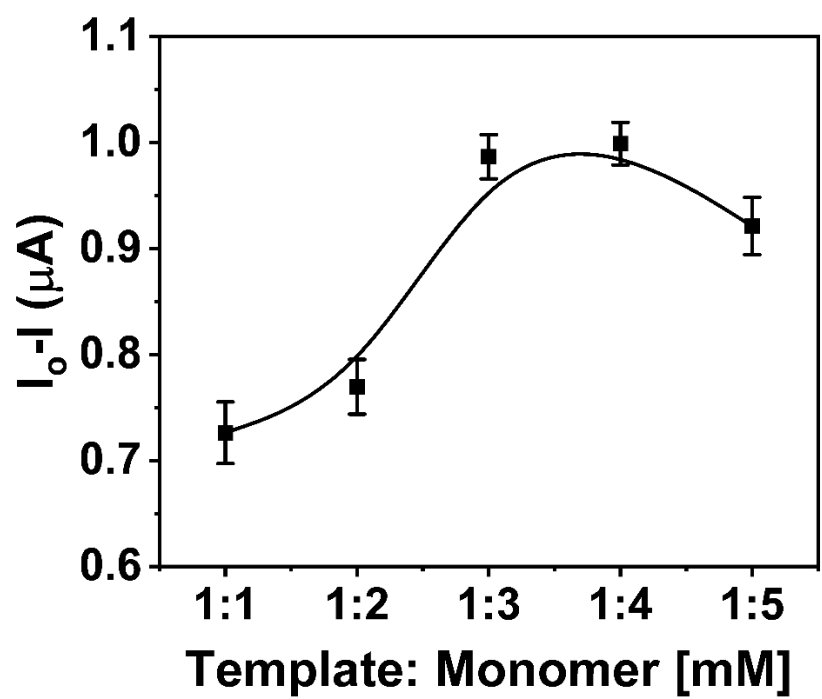

**Figure- S9:** Effect of the template: monomer (cortisol: pyrrole) at different ration starting from 1:1, 1:2, 1:3, 1:4, 1:5 mM in the presence of cortisol (1.0 pg mL$^{-1}$) in binding buffer (0.10 M, pH, 7.4).

**Supplementary Figure S10**

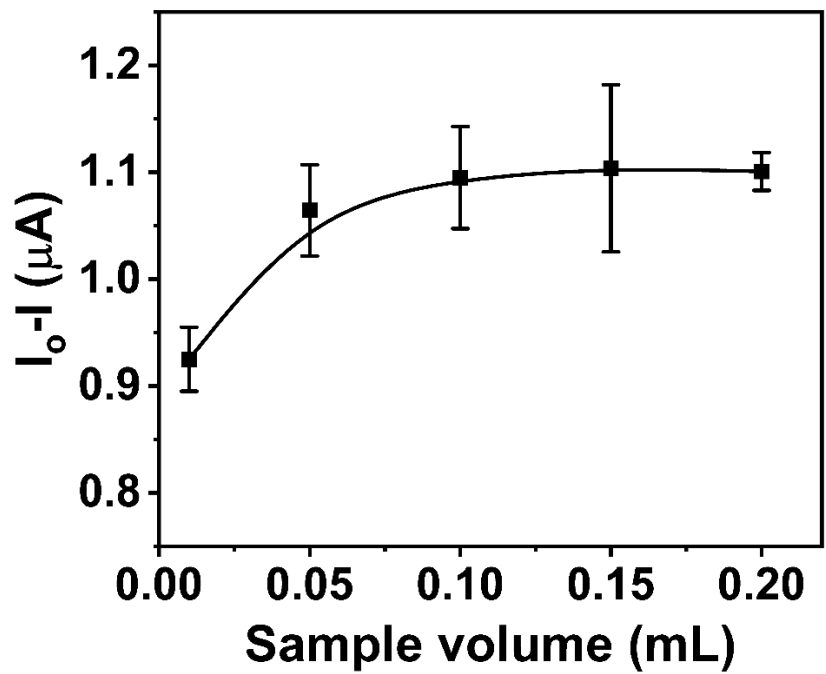

**Figure- S10:** Effect of sample volume (0.01, 0.05, 0.10, 0.15, and 0.20 mL) in the presence of cortisol (1.0 pg mL$^{-1}$) in binding buffer (0.10 M, pH, 7.4).

**Supplementary Figure S11**

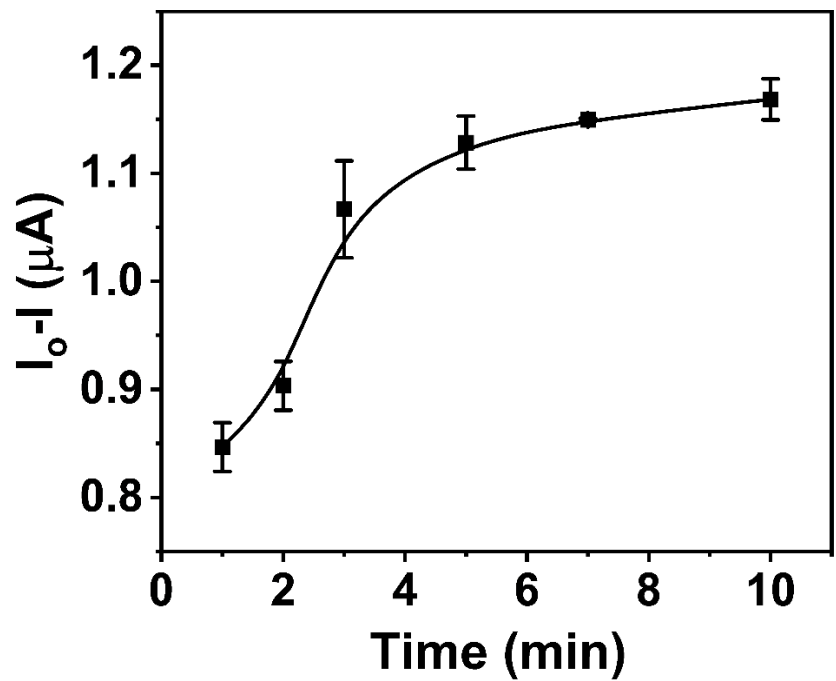

**Figure- S11:** Effect of incubation time (1, 2, 3, 5, 7, 10 minutes) on sensor performance in the presence of cortisol (1.0 pg mL$^{-1}$) in binding buffer (0.10 M, pH, 7.4).

**Supplementary Figure S12**

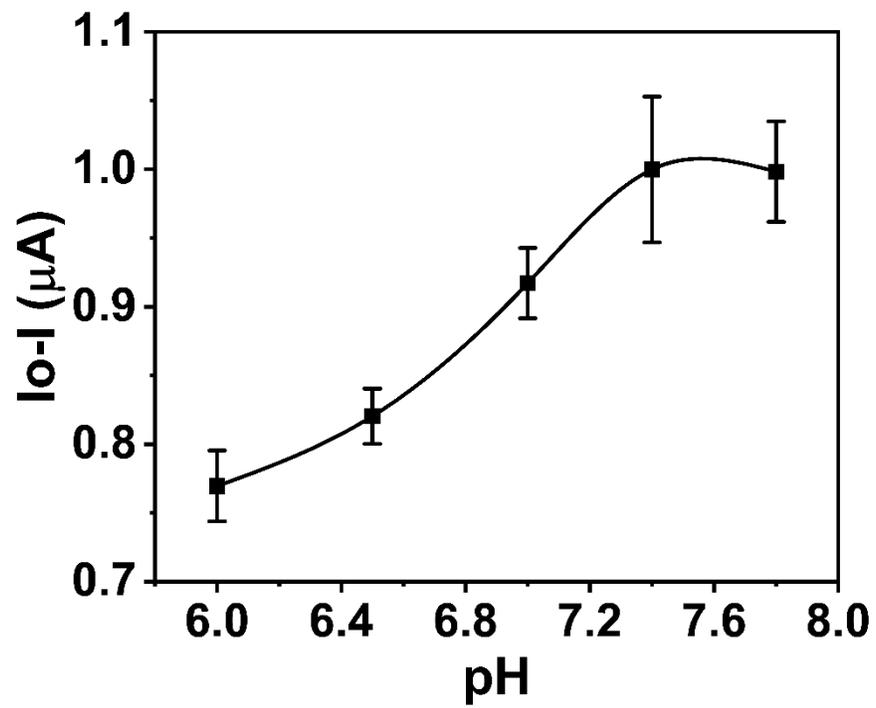

**Figure- S12:** Effect of pH (6.0, 6.5, 7.0, 7.4, 7.8) on sensor performance in the presence of cortisol (1.0 pg mL$^{-1}$) in binding buffer (0.10 M).

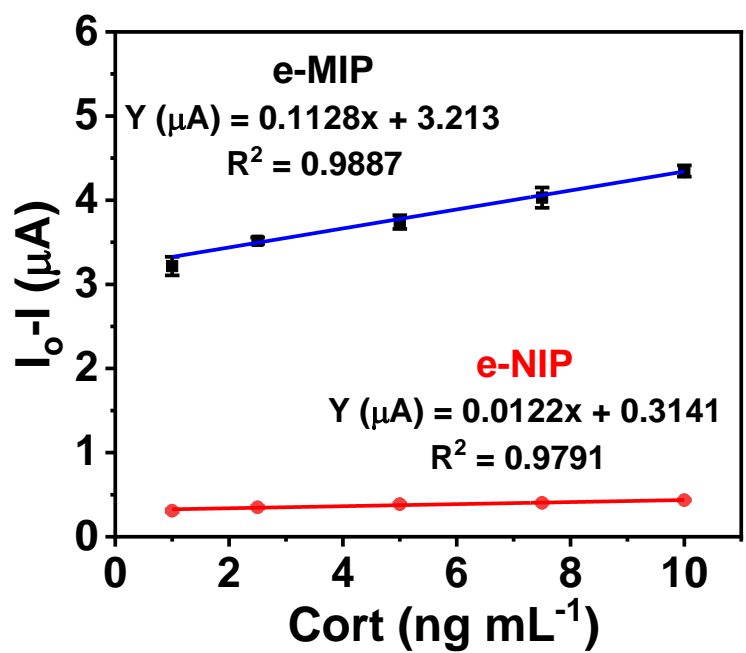

**Figure S13**: Calibration curve of cort-eMIP and cort-eNIP sensor in PBS (0.10 M, pH 7.4) with cortisol ranges from 1.0 to 10 ng mL$^{-1}$.

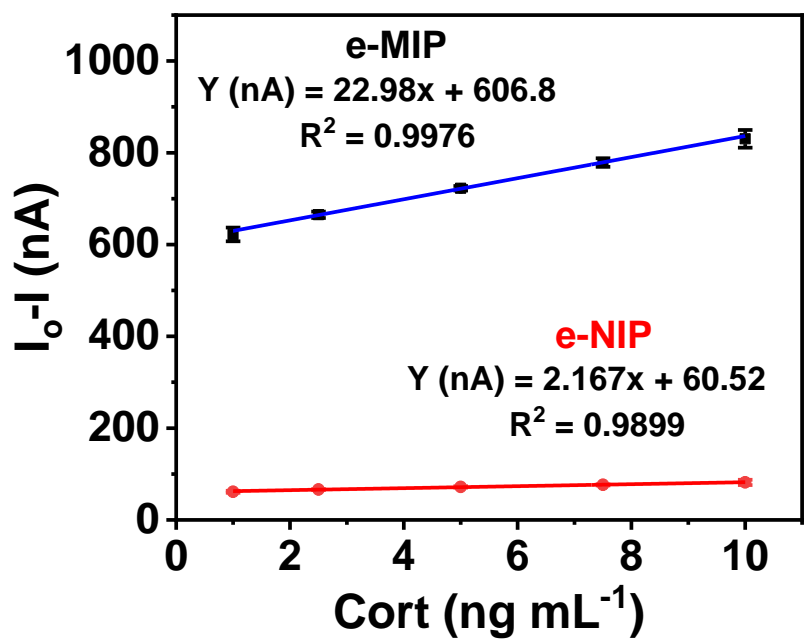

**Figure S14**: Calibration curve of cort-eMIP and cort-eNIP sensor in artificial saliva sample with cortisol ranges from 1.0 to 10 ng mL$^{-1}$.

**Supplementary Table 1 (ST1).** Comparison of analytical performance between the developed saliva-floss platform and other saliva diagnostics for cortisol detection

| S. No. | Techniques/ platform | Linear range | LOD | Reference |
|---|---|---|---|---|
| 1. | SPR-based immunosensor | 1- 100 ng mL$^{-1}$ | 1.0 ng mL$^{-1}$ | [3] |
| 2. | Molecularly imprinted polymer-modified organic FET | 0-4 ng mL$^{-1}$ | 0.72 ng mL$^{-1}$ | [4] |
| 3. | Aptasensor | 0.5–15 ng mL$^{-1}$ | 0.37 ng mL$^{-1}$ | [5] |
| 4. | Lateral flow assay (immunosensor) | 0.0 – 50 ng mL$^{-1}$ | 2.5 ng mL$^{-1}$ | [6] |
| 5. | Organic field-effect transistor (OFET) functionalized with MIP | 0.0- 40.0 μg L$^{-1}$ | 0.72 μg L$^{-1}$ (or 2.0 nM) | [7] |
| 6. | Molecularly Imprinted Polymer Electrochemical Sensor | 0.1 to 100 nM | 0.036 nM | [8] |
| 7. | Saliva-sensing floss (cort-eMIP/PPy/PB/PLEG-sensor) | 0.10 to 10000 pg mL$^{-1}$ | 0.023 pg mL$^{-1}$ (artificial saliva)<br><br>0.048 pg mL$^{-1}$ (human saliva) | **This work** |

**Supplementary Table 2 (ST2)**. Recovery studies of cortisol detection in spiked human saliva sample (n = 3)

| S. No. | Cortisol concentration added (pg mL$^{-1}$) | Cortisol concentration found (pg mL$^{-1}$) | % recovery | % RSD |
|---|---|---|---|---|
| 1. | 10.0 | 9.898 ± 0.120 | 98.98 | 1.21 |
| 2. | 100.0 | 98.64 ± 1.53 | 98.64 | 1.55 |
| 3. | 1000.0 | 1020.4 ± 14.30 | 102.04 | 1.40 |
| 4. | 10000.0 | 10029 ± 28.90 | 100.29 | 0.288 |

**\* % RSD- Percent relative standard deviation**